\documentclass{statsoc}
\usepackage{graphicx}
\usepackage{natbib}
\usepackage{delarray}
\usepackage{hyperref}
\usepackage{rotating}
\usepackage{url}
\usepackage{colortbl,color}
\usepackage{anysize}
\usepackage{multirow,amssymb,amsmath, amsfonts}

\def\singlespace{\def\baselinestretch{1}\@normalsize}

\def\boxit#1{\vbox{\hrule\hbox{\vrule\kern6pt\vbox{\kern6pt#1\kern6pt}\kern6pt\vrule}\hrule}}

\renewcommand{\baselinestretch}{1.3}
\newtheorem{theorem}{Theorem}

\newtheorem{lemma}{Lemma}
\newtheorem{proposition}{Proposition}

\newtheorem{assumption}{Assumption}

\newcommand{\bb}{\mbox{\bf b}}
\newcommand{\bff}{\mbox{\bf f}}
\newcommand{\bx}{\mbox{\bf x}}
\newcommand{\by}{\mbox{\bf X}}
\newcommand{\bA}{\mbox{\bf A}}
\newcommand{\ba}{\mbox{\bf a}}
\newcommand{\bw}{\mathrm{\bf W}}
\newcommand{\bu}{\mbox{\bf u}}

\newcommand{\bB}{\mbox{\bf B}}

\newcommand{\bD}{\mbox{\bf D}}
\newcommand{\bE}{\mbox{\bf E}}
\newcommand{\bF}{\mbox{\bf F}}

\newcommand{\bH}{\mbox{\bf H}}
\newcommand{\bI}{\mbox{\bf I}}
\newcommand{\bK}{\mbox{\bf K}}

\newcommand{\bQ}{\mbox{\bf Q}}

\newcommand{\bT}{\mbox{\bf T}}

\newcommand{\bV}{\mbox{\bf V}}
\newcommand{\bX}{\mbox{\bf X}}
\newcommand{\bW}{\mbox{\bf W}}
\newcommand{\bY}{\mbox{\bf Y}}
\newcommand{\bZ}{\mbox{\bf Z}}
\newcommand{\bone}{\mbox{\bf 1}}

\newcommand{\bmu}{\mbox{\boldmath $\mu$}}

\newcommand{\bgamma}{\mbox{\boldmath $\gamma$}}

\newcommand{\bGamma}{\mbox{\boldmath $\Gamma$}}

\newcommand{\bSigma}{\mbox{\boldmath $\Sigma$}}

\newcommand{\bOmega}{\mbox{\boldmath $\Omega$}}

\newcommand{\FDP}{\mbox{FDP}}
\newcommand{\FNR}{\mbox{FNR}}

\newcommand{\Var}{\mbox{Var}}
\newcommand{\Cov}{\mbox{Cov}}

\renewcommand{\textit}{}

\newcommand{\hw}{\widehat \bw}

\newcommand{\cov}{\mathrm{cov}}

\newcommand{\tr}{\mathrm{tr}}
\newcommand{\diag}{\mathrm{diag}}

\newcommand{\argmax}{\mathrm{argmax}}

\newcommand{\var}{\mathrm{var}}
\def\cov{\mbox{cov}}

\renewcommand{\hat}{\widehat}

\def\bB{\mbox{\bf B}}

\title[FDP under Unknown Dependence]{
Estimation of false discovery proportion with unknown dependence
\footnotetext{{\textit Address for correspondence:} Jianqing Fan, Department of Operations Research \& Financial Engineering, Princeton University,  Princeton, NJ 08544, USA. Email: jqfan@princeton.edu.}}
\author[J. Fan and X. Han]{Jianqing Fan$^{[1]}$ and Xu Han$^{[2]}$}
\address{$^{[1]}$Department of Operations Research $\&$ Financial
Engineering, Princeton University, Princeton,
New Jersey 08544, U.S.A. and School of Data Science, Fudan University, Shanghai, China}
\address{$^{[2]}$Department of
Statistics, Fox Business School, Temple University,
Philadelphia, Pennsylvania 19122, U.S.A.
}

\begin{document}
\begin{abstract}
Large-scale multiple testing with correlated test statistics arises frequently in many scientific research.  Incorporating correlation information in approximating false discovery proportion has attracted increasing attention in recent years. When the covariance matrix of test statistics is known, Fan, Han \& Gu (2012) provided an accurate approximation of False Discovery Proportion (FDP) under arbitrary dependence structure and some sparsity assumption. However, the covariance matrix is often unknown in many applications and such dependence information has to be estimated before approximating FDP. The estimation accuracy can greatly affect FDP approximation. In the current paper, we aim to theoretically study the impact of unknown dependence on the testing procedure and establish a general framework such that FDP can be well approximated. The impacts of unknown dependence on approximating FDP are in the following two major aspects:  through estimating eigenvalues/eigenvectors and through estimating marginal variances. To address the challenges in these two aspects,  we firstly develop general requirements on estimates of eigenvalues and eigenvectors for a good approximation of FDP. We then give conditions on the structures of covariance matrices that satisfy such requirements. Such dependence structures include banded/sparse covariance matrices and (conditional) sparse precision matrices. Within this framework, we also consider a special example to illustrate our method where data are sampled from an approximate factor model, which encompasses most practical situations. We provide a good approximation of FDP via exploiting this specific dependence structure. The results are further generalized to the situation where the multivariate normality assumption is relaxed. Our results are demonstrated by simulation studies and some real data applications.

\keywords{Large-scale multiple testing, dependent test statistics, false discovery proportion, unknown covariance matrix, approximate factor model}
\end{abstract}



\section{Introduction}
The correlation effect of dependent test statistics in large-scale multiple testing has attracted considerable attention in recent years. In microarray experiments, thousands of gene expressions are usually correlated when cells are treated. Applying standard Benjamini \& Hochberg (1995, B-H) or Storey (2002)'s procedures for independent test statistics can lead to inaccurate false discovery control. Statisticians have now reached the conclusion that it is important and necessary to incorporate the dependence information in the multiple testing procedure. See Efron (2007, 2010), Leek \& Storey (2008), Schwartzman \& Lin (2011) and Fan, Han \& Gu (2012).

Consideration of multiple testing procedure for dependent test statistics dates back to early 2000's. Benjamini \& Yekutieli (2001) proved that the false discovery rate can be controlled by the B-H procedure when the test statistics satisfy positive regression dependence on subsets (PRDS). Extension to a generalized stepwise procedure under PRDS has been proved by Sarkar (2002). Later Storey, et al. (2004) also showed that Storey's procedure can control FDR under weak dependence. Sun \& Cai (2009) developed a procedure where parameters underlying test statistics follow a hidden Markov model. Insightful results of validation for standard multiple testing procedures under more general dependence structures have been shown in Clarke \& Hall (2009). However, even if these procedures are valid under these special dependence structures, they still suffer from efficiency loss without considering the actual dependence information. In other words, there are universal upper bounds for a given class of covariance matrices.

A challenging question is how to incorporate the correlation effect in the testing procedure.  Efron (2007, 2010) in his seminal work obtained repeated test statistics based on the bootstrap sample from the original raw data,  took out the first eigenvector of the covariance matrix of the test statistics such that the correlation effect could be explained by a dispersion variate $A$, and estimated $A$ from the data to construct an estimate for realized FDP. Friguet, Kloareg \& Causeur (2009) and Desai \& Storey (2012) assumed that the data come directly from a strict factor model with independent idiosyncratic errors, and used the EM algorithms to estimate the number of factors, the factor loadings and the realized factors in the model and obtained an estimator for FDP by subtracting out realized common factors. The drawbacks of the aforementioned procedures are, however, restricted model assumptions and  the lack of formal justification.

Fan, Han \& Gu (2012) considered a general set-up for approximating FDP. They assumed that the test statistics are from a multivariate normal distribution with a known but arbitrary covariance matrix. Their idea is to apply spectral decomposition to the covariance matrix of test statistics and to use principal factors to account for dependency. This method is called Principal Factor Approximation (PFA). Under some sparsity assumption, the authors provided an accurate approximation of false discovery proportion (FDP) based on the eigenvalues and eigenvectors of the known covariance matrix.

A major restriction of the setup in Fan, Han \& Gu (2012) is that the covariance matrix of test statistics is known. Although the authors provided an interesting application with known covariance matrix, in many other cases, this matrix is usually unknown. For example, in microarray experiments, scientists are interested in testing if genes are differently expressed under different experimental conditions (e.g. treatments, or groups of patients). The dependence of test statistics is unknown in such applications. The problem of unknown dependence has at least two fundamental differences from the setting with known dependence: (a) Impact through estimating marginal variances. When the population marginal variances of the observable random variables are unknown, they have to be estimated first for standardization. In such a case, the popular choice of the test statistics will have $t$ distribution with dependence rather than the multivariate normal distribution considered in Fan, Han \& Gu (2012); (b) Impact through estimating eigenvalues/eigenvectors. Even if the population marginal variances of the observable random variables are known, estimation of eigenvalues/eigenvector can still significantly affect the FDP approximation. In various situations, FDP approximation can have inferior performance even if a researcher chooses the ``best" estimator for the unknown matrix. Therefore, more theoretical and methodological modifications are needed before directly applying PFA to unknown dependence setting.

The current paper aims to theoretically study the impact of unknown dependence on the testing procedure and establish a general framework for FDP approximation.  For the independence case,  this quantity depends asymptotically only on the number of true nulls.  For general case, as to be elucidated in Section 2.2 [around equation \eqref{eq6}], it is far more complicated, depending on the whole set of the unknown true nulls. Therefore, consistently estimating FDP is a hopeless task unless the signals are sparse. Under some sparsity assumption, FDP can be conservatively estimated by taking the null proportion to be one.  But this will cause other technical problems.  Instead, we will focus on a statistical quantity $\FDP_A$ (see equation (\ref{eq6})) and estimate it directly. $\FDP_A$ can be viewed as an asymptotic upper bound of FDP, and correspondingly the expectation of $\FDP_A$ is the asymptotic upper bound of the conventional FDR. For the challenges from the unknown dependence, since the impact of aspect (b) is even more important than that of aspect (a), we will first develop requirements for estimated eigenvalues and eigenvectors. Surprisingly, for a good estimate of this upper bound, we do not need these estimates of eigenvalues and eigenvectors to be consistent themselves. This finding relaxes the consistency restriction of covariance matrix  estimation under operator norm. Our framework of FDP approximation encompasses both weak dependence and strong dependence, including banded matrices, (conditional) sparse matrices, (conditional) sparse precision matrices, etc.

As a specific example,  we will consider the covariance matrices with an approximate factor structure. This factor model encompasses a majority of statistical applications and is a generalization to the model in Friguet, Kloareg \& Causeur (2009) and Desai \& Storey (2012). After applying Principal Orthogonal complEment Thresholding (POET) estimators (Fan, Liao \& Mincheva, 2013) to estimate the unknown covariance matrix, we can then assess FDP. This combination of POET  to estimate the covariance matrix and PFA to approximate FDP should be applicable to most practical situations and is the method that we recommend for practice.

We will also examine the impact of unknown marginal variances and generalize our results to the situation when the test statistics have $t$ distribution with dependence, which is beyond the multivariate normal assumption. This dependent $t$ distribution is not the conventional multivariate $t$ distribution. We will show that our proposed method is still applicable to this more general situation. The performance of our procedure is further evaluated by simulation studies and real data analysis.

The organization of the rest of the  paper is as follows: Section 2 provides background information of large scale multiple testing under dependency and Principal Factor Approximation (PFA), Section 3 includes the theoretical study on FDP approximations, Section 4 contains simulation studies, and Section 5 illustrates the methodology via an application to a microarray data set.  Throughout this paper, we use $\lambda_{\min}(\bA)$ and $\lambda_{\max}(\bA)$ to denote the minimum and maximum eigenvalues of a symmetric matrix $\bA$. We also denote the Frobenius norm $\|\bA\|_F=tr^{1/2}(\bA^T\bA)$, the operator norm $\|\bA\|=\lambda_{\max}^{1/2}(\bA^T\bA)$, and the induced norms $\|\bA\|_1=\max_{1\leq j\leq p}\sum_{i=1}^p|a_{ij}|$ and $\|\bA\|_{\infty}=\max_{1\leq i\leq p}\sum_{j=1}^p|a_{ij}|$.

The proposed method POET-PFA can be easily implemented by the R package ``pfa" (version 1.1) on https://cran.r-project.org. The simulation codes and the data set can be found in the supplementary materials.

\section{Approximation of FDP}
Suppose that the observed data $\{\bX_i\}_{i=1}^n$ are $p$-dimensional independent random vectors with $\bX_i\sim N_p(\bmu,\bSigma)$.  The mean vector $\bmu=(\mu_1,\cdots,\mu_p)^T$ is a high dimensional sparse vector, but we do not know which ones are the nonvanishing signals. Let $p_0 = \#\{j: \mu_j=0\}$ and $p_1 = \#\{j: \mu_j\neq0\}$ so that $p_0+p_1=p$. 
We wish to test which coordinates of $\bmu$ are signals based on the realizations $\{\bx_i\}_{i=1}^n$.

Consider the test statistics $\bZ=\sqrt{n}\overline{\bX}$ in which $\overline{\bX}$ is the sample mean of $\{\bX_i\}_{i=1}^n$. Then, $\bZ\sim N_p(\sqrt{n}\bmu,\bSigma)$. Standardizing the test statistics $\bZ$, we assume for simplicity that $\bSigma$ is a correlation matrix. Let $\bmu^{\star}=(\mu_1^{\star},\cdots,\mu_p^{\star})^T=\sqrt{n}\bmu$. Then, multiple testing $H_{0j}: \mu_j=0$ vs $H_{1j}: \mu_j\neq0$ is equivalent to test $H_{0j}: \mu_j^{\star}=0$ vs $H_{1j}: \mu_j^{\star}\neq0$ based on the test statistics $\bZ=(Z_1,\cdots,Z_p)^T$. The P-value for the $j^{th}$ hypothesis is $2\Phi(-|Z_j|)$, where $\Phi(\cdot)$ is the cumulative distribution function of the standard normal distribution.  We use a threshold value $t$ to reject the hypotheses which have p-values smaller than $t$. Define $R(t) =\#\{P_j: P_j\leq t\}$ as the number of discoveries and $V(t)=\#\{\text{true null}: P_j\leq t\}$ the number of false discoveries $V(t)$, where $P_j$ is the p-value for testing the $j$th hypothesis.  Our interest focuses on approximating the false discovery proportion $\FDP(t)=V(t)/R(t)$, here and hereafter the convention $0/0=0$ is always used.  Note that $R(t)$ is observable, and $\FDP(t)$ is a realized but unobservable random variable. 
In comparison with FDR(t) = $E [\FDP(t)$], an average of FDP for hypothetical replications of experiments, FDP concerns about the number of false discoveries given the experiment.


The normality assumption is idealization. 
In the current paper, we will show both theoretically and numerically that even if the normality assumption is violated, our results are still applicable for a more general setting.

\subsection{Impact of dependence on the false discoveries}
The number of false discoveries $V(t)$ is an important quantity in multiple testing. It is a realized but unobservable value for a given experiment.
To gain the insight on how the dependence of test statistics impacts on the number of false discoveries, let us first illustrate this by a simple example:  The test statistic depend on a common unobservable factor $W$ in the following model
\begin{equation} \label{eq1}
Z_i = \mu_i^\star + b_i W + (1 - b_i^2)^{1/2} \varepsilon_i \sim N(\mu_i^\star , 1), 
\end{equation}
where $W$ and $\{\varepsilon_i\}_{i=1}^n$ are independent, having the standard normal distribution.
Let $z_\alpha$ be the $\alpha$-quantile of the standard normal distribution and ${\cal N} = \{i: \mu_i^\star = 0\}$ is the true null set.  Then,
\begin{equation*}
V(t) = \sum_{i \in {\cal N}} I(|Z_i| > -z_{t/2}) = \sum_{i \in {\cal N}} \Big[I\big(  \varepsilon_i > a_i(-z_{t/2} -  b_i W )\big) +I\big(  \varepsilon_i < a_i(z_{t/2} -  b_i W )\big)\Big],
\end{equation*}
where $a_i = (1 - b_i^2)^{-1/2}$.
By using the law of large numbers, conditioning on $W$, under some mild conditions, we have
\begin{equation} \label{eq2}
p_0^{-1}V(t) =p_0^{-1} \sum_{i \in {\cal N}} [\Phi(a_i (z_{t/2} + b_i W)) + \Phi(a_i (z_{t/2} - b_i W))]+o_p(1).
\end{equation}
The dependence of $V(t)$ on the realization $W$ is evidenced in (\ref{eq2}).  For example, if $b_i = \rho$,
\begin{equation} \label{eq3}
p_0^{-1}V(t) = \left [\Phi \left ( \frac{z_{t/2} + \rho W}{\sqrt{1 - \rho^2}} \right ) + \Phi \left ( \frac{z_{t/2} - \rho W}{\sqrt{1 - \rho^2}} \right ) \right ]+o_p(1).
\end{equation}
When $\rho = 0$, $p_0^{-1}V(t) \approx t$ as expected.  To quantify the dependence on the realization of $W$, let $p_0 = 1000$ and $t = 0.01$ and $\rho = 0.8$ so that
$$
p_0^{-1}V(t) \approx [\Phi((-2.236 + 0.8 W)/0.6) +  \Phi((-2.236 - 0.8 W)/0.6)].
$$
When $W = -3, -2, -1, 0$, the values of $p_0^{-1}V(t)$ are approximately  0.608, 0.145, 0.008 and 0, respectively, which depends heavily on the realization of $W$.  This is in contrast with the independence case in which $p_0^{-1}V(t)$ is always approximately 0.01.

Despite the dependence of $V(t)$ on the realized random variable $W$, the common factor can be inferred from the observed test statistics.  For example, ignoring sparse $\mu_i^\star$ in (\ref{eq1}), we can estimate $W$ via the simple least-squares: Minimizing $\sum_{i=1}^p (Z_i - b_i W)^2$ with respect to $W$. Substituting the estimate into (\ref{eq3}) and replacing  $p_0$ by $p$, or more generally substituting the estimate into (\ref{eq2}) and replace $\cal N$ by the entire set, we obtain an estimate of $V(t)$ under dependence.  A robust implementation is to use $L_1$-regression which finds $W$ to minimize $\sum_{i=1}^p |Z_i - b_i W|$ or to use penalized least-squares such as $\sum_{i=1}^p (Z_i - \mu_i - b_i W)^2 + \lambda \sum_{i=1}^p |\mu_i|$ to explore the sparsity of $\mu$.
This is the basic idea behind Fan, Han \& Gu (2012).

\subsection{Principal Factor Approximation}
The Principal Factor Approximation, introduced by Fan, Han \& Gu (2012), is a generalization of the idea in Section 2.1. Let $\lambda_1,\cdots,\lambda_p$ be the eigenvalues of correlation matrix $\bSigma$ in non-increasing order, and $\bgamma_1,\cdots,\bgamma_p$ be their corresponding eigenvectors. For a given integer $k$,  decompose $\bSigma$  as
\begin{equation*}
\bSigma=\bB\bB^T+\bA,
\end{equation*}
where $\bB=(\sqrt{\lambda_1}\bgamma_1,\cdots,\sqrt{\lambda_k}\bgamma_k)$ are unnormalized first $k$ principal components and $\bA=\sum_{i=k+1}^p\lambda_i\bgamma_i\bgamma_i^T$. Correspondingly, decompose the test statistics $\bZ\sim N(\bmu^{\star},
\bSigma)$ stochastically as
\begin{equation}\label{eq5}
\bZ =\bmu^{\star}+\bB\bW +\bK,
\end{equation}
where $\bW \sim N_k(0,\bI_k)$ are $k$ common factors and  $\bK \sim N(0,\bA)$ are the errors, independent of $\bW$. Define the oracle $\FDP(t)$ as
\begin{equation}\label{j12}
\FDP_{oracle}(t)=\sum_{i\in\{true \ nulls\}}[\Phi(a_i(z_{t/2}+\eta_i))+\Phi(a_i(z_{t/2}-\eta_i))]/R(t)
\end{equation}
where $a_i=(1-\|\bb_i\|^2)^{-1/2}$, $\eta_i=\bb_i^T\bW$ and $\bb_i^T$ is the $i^{th}$ row of $\bB$.  This is clearly a generalization of (\ref{eq2}). Then, an examination of the proof of Fan, Han \& Gu (2012) yields the following result:

\begin{proposition}
If   (C0): \ \  $p^{-1} \sqrt{\lambda_{k+1}^2+\cdots+\lambda_p^2} =O(p^{-\delta})$ for some  $\delta>0$, then
on the event $\{p^{-1} R(t) > c p^{-\theta}\}$ for some $c> 0$ and $\theta\geq0$, we have $|\FDP_{oracle}(t)-\FDP(t)|=O_p(p^{-(\delta/2-\theta)})$.
\end{proposition}

The above proposition was established in the proof of Theorem 1 of Fan, Han \& Gu (2012) 
under (C0) and the assumption that $\theta = 0$. Here we
allow $\theta > 0$ and $R(t)$ can stochastically grow slower than $p$.
Suppose we choose $k'>k$. Then by (C0) it is easy to see that the associated convergence rate is no slower than $p^{-(\delta/2-\theta)}$. This explains that with more common factors in model (\ref{eq5}), $|\FDP_{oracle}(t)-\FDP(t)|$ converges to zero faster as $p\rightarrow\infty$. This result will be useful for the discussion about determining number of factors in section 3.1. Condition (C0) in Proposition 1 implies that if $\|\bSigma\| = o(p^{1/2})$, we can take $k=0$. In other words, $\|\bSigma\| = o(p^{1/2})$ can be regarded as the condition for weak dependence of multiple testing problem.   For the mean-square convergence of $V(t)$, see Azriel and Schwartzman (2015).

Since we do not know which coordinates of $\bmu$ vanish, $\FDP_{oracle}(t)$ can be approximated by
\begin{equation}\label{eq6}
\FDP_A(t)=\sum_{i=1}^p[\Phi(a_i(z_{t/2}+\eta_i))+\Phi(a_i(z_{t/2}-\eta_i))]/R(t).
\end{equation}
This provides a useful upper bound for estimating $\FDP(t)$. For the independence case, in which $a_i=1$ and $\|\bb_i\| = 0$, $\FDP_{oracle} (t) = p_0 t / R(t)$.  It can be consistently estimated by estimating one parameter $p_0$.  For dependence case, however, we need to know the whole set of ``true null'' and this is an impossible task.  Therefore the upper bound becomes an estimable statistical quantity that is frequently used in practice.



The principal factor approximation (PFA) method of Fan, Han \& Gu (2012) is to define
\begin{equation}\label{eq7}
\widehat{\FDP}_{A}(t)=\sum_{i=1}^p[\Phi(a_i(z_{t/2}+
\widetilde{\eta}_i))+\Phi(a_i(z_{t/2}-\widetilde{\eta}_i))]/R(t),
\end{equation}
where $\widetilde{\eta}_i=\bb_i^T\widehat{\bw}$ for an estimator $\widehat{\bw}$ of $\bw$. Then, under mild conditions, Fan, Han \& Gu (2012) shows $\big|\widehat{\FDP}_{A}(t)-\FDP_A(t)\big|=O_p(\|\widehat{\bw}-\bw\|)$.

For the estimation of $\bw$, since $\bmu^{\star}$ is sparse, one can consider the following penalized least-squares estimator based on model (\ref{eq5}). Namely, $\widehat{\bw}$ is obtained by minimizing
\begin{equation}\label{eq8}
\sum_{i=1}^p(z_i-\mu_i^{\star}-\bb_i^T\bw)^2+\sum_{i=1}^pp_{\lambda}(|\mu_i^{\star}|)
\end{equation}
with respect to $\bmu^{\star}$ and $\bw$, where $p_{\lambda}$ can be the $L_1$ or SCAD penalty function.
When  $p_{\lambda}(|\mu_i^{\star}|)=\lambda|\mu_i^{\star}|$, the optimization problem in (\ref{eq8}) is equivalent to
\begin{equation}\label{eq9}
\min_{\bw}\sum_{i=1}^p\psi(z_i-\bb_i^T\bw)
\end{equation}
where $\psi(\cdot)$ is the Huber loss function (Fan, Tang \& Shi, 2012). In Fan, Han \& Gu (2012), they also considered an alternative loss function for (\ref{eq9}), the least absolute deviation loss:
\begin{equation} \label{eq9a}
        \min_{\bw}\sum_{i=1}^p |z_i-\bb_i^T\bw|.
\end{equation}
Fan, Tang \& Shi (2012) studies (\ref{eq8}) rigourously.
They show that the penalized estimator of $\bw$ is consistent and that its asymptotic distributions are Gaussian. 

\subsection{PFA with Unknown Covariance}
The $\widehat{\FDP}_{A}(t)$ in (\ref{eq7}) is based on eigenvalues $\{\lambda_i\}_{i=1}^k$ and eigenvectors $\{\bgamma_i\}_{i=1}^k$ of the true covariance matrix $\bSigma$.  When $\bSigma$ is unknown, we need an estimate $\widehat{\bSigma}$. Let $\widehat{\lambda}_1,\cdots,\widehat{\lambda}_p$ be eigenvalues of $\widehat{\bSigma}$ in a non-increasing order and  $\widehat{\bgamma}_1,\cdots,\widehat{\bgamma}_p\in\mathbb{R}^p$ be their corresponding eigenvectors.   One can obtain an approximation of FDP by substituting unknown  eigenvalues and eigenvectors in (\ref{eq7}) by their corresponding estimates.  Two questions arise naturally:
\begin{itemize}
\item[(1)] What are the requirements for the estimates of $\{\lambda_i\}_{i=1}^k$ and $\{\bgamma_i\}_{i=1}^k$ such that $\big|\widehat{\FDP}_{A}(t)-\FDP_A(t)\big| = o_p(1)$?
\item[(2)] Under what dependence structures of $\bSigma$, can such estimates of $\{\lambda_i\}_{i=1}^k$ and $\{\bgamma_i\}_{i=1}^k$ be constructed?
\end{itemize}
The current paper will address these two questions.

\section{Main Result}

We first present the results for a generic estimator $\hat{\bSigma}$, and then consider a special example in this general framework, approximate factor model, to illustrate the impact of unknown dependence on the testing procedure.

\subsection{Required Accuracy}
Suppose that $(C0)$ is satisfied for $\bSigma$. Let $\widehat{\bSigma}$ be an estimator of $\bSigma$, and correspondingly we have $\{\widehat{\lambda}_i\}_{i=1}^k$ and $\{\widehat{\bgamma}_i\}_{i=1}^k$ to estimate $\{\lambda_i\}_{i=1}^k$ and $\{\bgamma_i\}_{i=1}^k$. Analogously, we define $\hat{\bB}$ and $\widehat{\bb}_i$.
Note that we only need to estimate the first $k$ eigenvalues and eigenvectors but not all of them. 

The realized common factors $\bW$ can be estimated robustly by using (\ref{eq8}) and (\ref{eq9}) with $\bb_i$ replaced by $\widehat{\bb}_i$.  To simplify the technical arguments, we  simply use the least-squares estimate
 \begin{equation} \label{eq10}
    \widehat{\bW}=(\widehat{\bB}^T\widehat{\bB})^{-1}\widehat{\bB}^T\bZ,
\end{equation}
which ignores the $\bmu^\star$ in (\ref{eq5}) and replaces $\bB$ by $\hat{\bB}$. Define
\begin{equation}\label{eq11}
\widehat{\FDP}_{U}(t)=\sum_{i=1}^p[\Phi(\widehat{a}_i(z_{t/2}+\widehat{\eta}_i))+\Phi(\widehat{a}_i(z_{t/2}-\widehat{\eta}_i))]/R(t)
\end{equation}
where $\widehat{a}_i=(1-\|\widehat{\bb}_i\|^2)^{-1/2}$ and $\widehat{\eta}_i=\widehat{\bb}_i^T\widehat{\bW}$. Then we have the following result.

\begin{theorem} \label{thm1}
On the event ${\cal E}$ that
\begin{itemize}
\item[(C1)] $R(t)^{-1}=O(p^{-(1-\theta)})$ for some $\theta\geq0$,
\item[(C2)] $\max_{i \leq k} \|\widehat{\bgamma}_i-\bgamma_i\|=O(p^{-\kappa})$ for $\kappa>0$,
\item[(C3)] $\sum_{i =1 }^ k |\widehat{\lambda}_i-\lambda_i|=O(p^{1-\nu})$ for $\nu>0$,
\item[(C4)] $\widehat{a}_i\leq\tau_1$ and $a_i\leq\tau_2$ $\forall i=1,\cdots,p$ for some finite constants $\tau_1$ and $\tau_2$,
\end{itemize}
we have
\begin{equation*}
|\widehat{\textrm{\FDP}}_{U}(t)-\textrm{\FDP}_A(t)|
=O_p\Big(p^{\theta}\big(p^{-\nu}+kp^{-\kappa}+\|\bmu^{\star}\|p^{-1/2}\big)\Big).
\end{equation*}
\end{theorem}
Note that FDP$(t)=V(t)/R(t)$ in which $R(t)$ is observable and known. Approximating FDP$(t)$ amounts to approximating $V(t)$, which does not rely on Condition (C1).  In high-dimensional application, $t$ can be chosen to slowly decrease with $p$, as in Donoho \& Jin (2004, 2006).  Our result on the approximation of $V(t)$ continues to hold for $t$ that depends on $p$, i.e. $t_p$.  If Condition (C1) holds for $t_p$, then Theorem 1 follows for $t_p$.

Using $\sum_{i=1}^k \lambda_i \leq \tr(\bSigma) = p$, we have
$\sum_{i =1 }^ k |\widehat{\lambda}_i-\lambda_i| \leq  p \max_{i \leq k} |\widehat{\lambda}_i/\lambda_i -1|$.
Thus, Condition (C3) holds with high probability when
$\max_{i \leq k} |\widehat{\lambda}_i/\lambda_i -1| = O_p(p^{-\nu})$.
The latter is particularly relevant when eigenvalues are spiked. The third term in the convergence result comes really from the least-squares estimate.  If a more sophisticated method such as (\ref{eq8}) or (\ref{eq9}) is used, the bias will be smaller (Fan, Tang \& Shi, 2012).  We do not plan to pursue along this line to facilitate the presentation.

In Theorem 1, we assume that the number of factors $k$ is known. When $k$ has to be estimated, we will apply the eigenvalue ratio (ER) estimator in Ahn \& Horenstein (2013). The ER estimator is defined as $\widehat{k}_{ER}=\argmax_{1\leq k\leq k_{\max}}(\widetilde{\lambda}_k/\widetilde{\lambda}_{k+1})$, where $\widetilde{\lambda}_i$ is the $i$th largest eigenvalue of the sample covariance matrix and $k_{\max}$ is the maximum possible number of factors. Under mild regularity conditions, this estimator has been shown consistent. Similar idea has also been adopted by Lam \& Yao (2012). Therefore, to simplify the presentation, we will use a known $k$ for the theoretical development in the current paper, but for the numerical studies in Section 4 and 5 we will apply the ER estimator for estimating $k$. An over estimate of $k$ does not do as much harm to approximating FDP, as long as the unobserved factors are estimated with reasonable accuracy.  This is due to the fact that Condition (C0) is also satisfied for a larger $k$ and will be verified via simulation.   On the other hand, an underestimate of $k$ can result in the approximated FDP with inferior performance, due to missing important factors to capture dependency.

\subsection{Impact of estimating marginal variances}
In the previous sections, we assume that $\bSigma$ is a correlation matrix. In practice, the marginal variances $\{\sigma_j^2\}$ are unknown and need to be estimated. These estimates are used to normalize the testing problem.   Suppose $\{\widehat{\sigma}_j^2\}_{j=1}^p$ are the diagonal elements of $\widehat{\bSigma}$, an estimate of $\bSigma$. Conditioning on $\{\widehat{\sigma}_j\}_{j=1}^p$, assume $\widehat{\bD}^{-1} \sqrt{n}\overline{\bX} \sim N(\sqrt{n}\widehat{ \bD}^{-1} \bmu, \widetilde{\bSigma}), \widetilde{\bSigma} =\widehat{ \bD}^{-1} \bSigma\widehat{\bD}^{-1}$, where $\widehat{\bD} =\diag\{\widehat{\sigma}_1, \cdots, \widehat{\sigma}_p\}.$ When $\hat{\bSigma}$ is the sample covariance matrix, it is well-known that $\hat{\bSigma}$ and $\overline{\bX}$ are independent and the aforementioned assumption holds.  Then $\widetilde{\bSigma}$ is approximately the same as the correlation matrix as long as $\{\widehat{\sigma}_j\}_{j=1}^p$ converges uniformly to $\{{\sigma}_j\}_{j=1}^p$.  Thanks to the Gaussian tails, this indeed holds for the sequence of the marginal sample covariances (Bickel \& Levina, 2008a). Our simulations show the small impact of estimating the marginal variances.

The unconditional distribution of $\widehat{\bD}^{-1}\sqrt{n}\overline{\bX}$ is not a multivariate normal. To address this issue, let $\overline{X}_{(j)}=n^{-1}\sum_{i=1}^nX_{ij}$ and $\widehat{\sigma}_j^2=(n-1)^{-1}\sum_{i=1}^n(X_{ij}-\overline{X}_{(j)})^2$ and consider the standardized test statistics $T_j=\sqrt{n}\overline{X}_{(j)}/\widehat{\sigma}_j$. Then, for the true nulls, each $T_j$ follows the $t_{n-1}$-distribution, and $(T_j, T_l)$ have a bivariate $t$ distribution. See Siddiqui(1967). However, $\{T_j\}_{j=1}^p$ do not follow the multivariate $t$ distribution introduced in Kotz \& Nadarajah (2004), because $\{\widehat{\sigma}_j\}_{j=1}^p$ are also dependent of each other through $\bSigma$.  Therefore, in the following presentation, we will call the joint distribution of $\{T_j\}_{j=1}^p$ a dependent $t$ distribution rather than a multivariate $t$ distribution to avoid any confusion. Let $F_{n-1}()$ denote the cumulative distribution function of a $t_{n-1}$ random variable, and let $q_{t/2}$ denote the $t/2$ quantile of $F_{n-1}$. The p-values are calculated as $P_j=2F_{n-1}(-|T_j|)$. We use threshold $t$ and reject the $j$th hypothesis if $P_j\leq t$.

Similar to the definition of $\widehat{\FDP}_U(t)$ in section 3.1, we use the least squares estimate
\begin{equation*}
\widehat{\bW}_G=(\widehat{\bB}^T\widehat{\bB})^{-1}\widehat{\bB}^T\bT,
\end{equation*}
where $\bT=(T_1,\cdots, T_p)^T$. Define $\widehat{\FDP}_{U,G}(t)=\sum_{i=1}^p[\Phi(\widehat{a}_i(z_{t/2}+\widehat{\eta}_{i,G}))+\Phi(\widehat{a}_i(z_{t/2}-\widehat{\eta}_{i,G}))]/R(t)$, where $\widehat{\eta}_{i,G}=\widehat{\bb}_i^T\widehat{\bW}_G$. In the above, $\widehat{\bB}$, $\widehat{\bb}_i$ and $\widehat{a}_i$ are calculated based on the estimated correlation matrix of $\bX$, and the subscript ``$G$" represents general covariance matrix $\bSigma$.

\begin{theorem}
Based on the test statistics $\{T_j\}_{j=1}^p$, suppose that the correlation matrix of $\bX$ satisfies condition (C0).   Then, on the event ${\cal E}$ in Theorem~\ref{thm1} , we have
\begin{equation*}
|\FDP_{oracle}(t)-\FDP(t)|=O_p\Big(p^{\theta}\big(p^{-\delta/2}+n^{-1/2}\big)\Big).
\end{equation*}
where $\FDP_{oracle}(t)$ is defined in (\ref{j12}) and
\begin{equation*}
|\widehat{\textrm{\FDP}}_{U,G}(t)-\textrm{\FDP}_A(t)|=O_p\Big(p^{\theta}\big(p^{-\nu}+kp^{-\kappa}+\|\bmu^{\star}\|p^{-1/2}+n^{-1/2}\big)\Big),
\end{equation*}
where $\FDP_A(t)$ is defined in (\ref{eq6}) corresponding to the correlation matrix of $\bX$.
\end{theorem}

The first result in Theorem 2 is similar to Proposition 1, except a term from the effect of the sample size $n$. This result suggests that under some mild conditions, we can still apply PFA method even if the effect of the marginal variance is considered.   Note that in the second result of Theorem 2, $\{\widehat{\lambda}_i\}$ and $\{\widehat{\bgamma}_i\}$ correspond to the estimated correlation matrix of $\bX$, and $\{\lambda_i\}$ and $\{\bgamma_i\}$ correspond to the population correlation matrix of $\bX$. This result is very similar to that established in Theorem 1. Therefore, to simplify the discussion and highlight the impact of estimator $\widehat{\bSigma}$ on the testing procedure, we will assume in the following sections 3.3--3.5 that the diagonal elements of $\bSigma$ are known and equal to 1. The simulation studies in section 4 are still based on the setup that $\bSigma$ has general and unknown diagonal elements.

Direct derivation of density function for the bivariate $t$ random variables is complicated and not useful for our proof. The proof of Theorem 2 is based on a Bayesian interpretation of bivariate $t$ distributions. The method is general and can be of independent interest for extending results under normality to dependent $t$ distributions.


\subsection{Results in Eigenvectors and Eigenvalues}
In Theorem 1, the convergence rate of $\widehat{\FDP}_U(t)$ critically depends on the estimated eigenvalues and eigenvectors. In the current section, we will study under what situations that conditions (C2) and (C3) can be satisfied.
\begin{lemma}  For any matrix $\hat{\bSigma}$, we have
\begin{equation*}
|\widehat{\lambda}_i-\lambda_i|\leq\|\widehat{\bSigma}-\bSigma\| \quad
\mbox{and} \quad \|\widehat{\bgamma}_i-\bgamma_i\|\leq\frac{\sqrt{2}\|\widehat{\bSigma}-\bSigma\|}
{\min(|\widehat{\lambda}_{i-1}-\lambda_i|,|\lambda_i-\widehat{\lambda}_{i+1}|)}.
\end{equation*}
\end{lemma}
The first result is referred to Weyl's Theorem (Horn \& Johnson, 1990) and the second result is called the $\sin\theta$ Theorem (Davis \& Kahan, 1970). They have been applied in sparse covariance matrix estimation (El Karoui, 2008;  Ma, 2013).  By Lemma 1, the consistency of eigenvectors and eigenvalues is directly associated with the operator norm consistency. Several papers have shown that under various conditions on $\bSigma$, $\widehat{\bSigma}$ can be constructed such that $\|\widehat{\bSigma}-\bSigma\|\rightarrow0$, which will be discussed in more details after the following Theorem 3.
\begin{theorem}
If $\lambda_i-\lambda_{i+1}\geq d_p$ for a sequence $d_p > 0$  for $i=1,\cdots,k$, then on the event ${\cal E}\cap \{ \|\widehat{\bSigma}-\bSigma\|=O(d_p p^{-\tau})\}$ for some $\tau>0$, for sufficiently large $p$,  we have
$$
    |\widehat{\FDP}_{U}(t)-\FDP_A(t)| = O_p\Big ( p^{\theta}\big(  k p^{-\tau} d_p/p +
     (k+1) p^{-\tau} + \|\bmu^{\star}\|p^{-1/2}\big) \Big ).
$$
\end{theorem}
Note that the first $k$ eigenvalues should be distinguished with a certain amount of gap
$d_p$.  The theorem is so written that it is applicable to both spike or non-spike case.  For the non-spike case, typically $d_p = d > 0$.  In this case, the covariance is estimated consistently and the first term in Theorem 3 now becomes $O_p\big ( kp^{-\tau -1 } \big )$.  For the spiked case such as the $k$-factor model (\ref{eq5}), the first $k$ eigenvalues are of order $p$ and the $(k+1)^{th}$ eigenvalue is of order 1 (Fan, Liao \& Mincheva, 2013).  Therefore, $d_p \asymp p$.  In this case, the covariance matrix can not be consistently estimated, and the first term is of order $O(kp^{-\tau})$.  See section 3.4 for additional details.

Depending on the structures of $\bSigma$ and different choices of $\widehat{\bSigma}$, we will have different requirements such that the event $\{\|\widehat{\bSigma}-\bSigma\|=O(d_pp^{-\tau})\}$ occurs with high probability. It is impossible for us to list all the references in the area of large covariance matrix estimation, but we will focus on several representative classes of $\bSigma$ structures and present relevant results.
\begin{itemize}
\item[1.] \textbf{Banded Matrix:} In Bickel \& Levina (2008a), the authors considered a class of banded matrices with decaying rate $\alpha$. After banding the sample covariance matrix, they constructed an estimator $\widehat{\bSigma}_1$, which has operator norm convergence rate as $\|\widehat{\bSigma}_1-\bSigma\|=O_p\Big(\big(\log p/n\big)^{\alpha/(2\alpha+2)}\Big)$.
\item[2.] \textbf{Sparse Matrix:} In Bickel \& Levina (2008b), a class of sparse covariance matrices is considered with sparsity parameters $c_0(p)$ and $q$ where $0\leq q\leq1$. With thresholding technique, they constructed an estimator $\widehat{\bSigma}_2$ which satisfies $\|\widehat{\bSigma}_2-\bSigma\|=O_p\Big(c_0(p)\big(\log p/n\big)^{(1-q)/2}\Big)$. In the special case when $q=0$ and $c_0(p)$ is bounded, this convergence rate is $(\log p/n)^{1/2}$.
\item[3.] \textbf{Sparse Precision Matrix:} 
    In Cai, Liu \& Luo (2011), they considered a class of sparse precision matrices $\bOmega=\bSigma^{-1}$ with sparsity parameters $s_0(p)$ and $q$. By a constrained $l_1$ minimization approach (CLIME), they constructed an estimator $\widehat{\bOmega}_3$ such that $\|\widehat{\bOmega}_3-\bOmega\|=O_p\Big(s_0(p)\big(\log p/n\big)^{(1-q)/2}\Big)$. Furthermore, for $\widehat{\bSigma}_3=(\widehat{\bOmega}_3)^{-1}$, under some mild conditions, it is easy to show that $\|\widehat{\bSigma}_3-\bSigma\|=O_p\Big(s_0(p)\big(\log p/n\big)^{(1-q)/2}\Big)$.
\end{itemize}

It is worth mentioning that the convergence rate of $\|\widehat{\bSigma}-\bSigma\|$ leading to some requirement of the sample size $n$. For example, in the special case of sparse matrix when $\|\widehat{\bSigma}-\bSigma\|=O_p((\log p/n)^{1/2})$, if it also satisfies the condition in Theorem 3 that $\|\widehat{\bSigma}-\bSigma\|=O_p(p^{-\tau})$, then the sample size $n$ has to be greater than $p^{2\tau}\log p$. This requirement of $n$ is of major importance in practice.

\subsection{Approximate Factor Model}
We will study the multiple testing problem where the test statistics have some strong dependence structure as a special example of Theorem 3. Assume the dependence of high-dimensional variable vector of interest can be captured by a few latent factors. This factor structure model has long history in financial econometrics (Engle \& Watson 1981, Bai 2003). It has also received considerable attention in genomic research (Friguet, Kloareg \& Causer 2009, Desai \& Storey 2012). Major restrictions in these models are that the idiosyncratic errors are independent. A more practicable extension is the approximate factor model (Chamberlain \& Rothschild 1983, Fan, Liao \& Mincheva, 2011, 2013).

The approximate factor model takes the form
\begin{equation}\label{eq12}
\by_i=\bmu+\bB\bff_i+\bu_i, \qquad i = 1, \cdots, n
\end{equation}
for each observation,
where $\bmu$ is a $p$-dimensional unknown sparse vector, $\bB=(\bb_1,\cdots,\bb_p)^T$
is the factor loading matrix, $\bff_i$ is a vector of common factors to the $i^{th}$ observations,
independent of the noise $\bu_i\sim N_p(0,\bSigma_u)$ where $\bSigma_u$ is
sparse.   The unobserved common factors $\bff_i$ drive the dependence of the measurements (e.g. gene expressions) within the $i^{th}$ sample.  Under model \eqref{eq12}, the covariance matrix of $\by_i$ is given by $\bSigma=\bB\cov(\bff)\bB^T+\bSigma_u$. We can also assume without loss of generality the identifiability condition: $\cov(\bff)=\bI_K$ and the columns of $\bB$ are orthogonal.  See Fan, Liao \& Mincheva (2013).

For the random errors $\bu$, let $\sigma_{u,ij}$ be the $(i,j)$th element of covariance matrix $\bSigma_u$ of $\bu$. Then we impose a sparsity condition on $\bSigma_u$:
\begin{equation}\label{j2}
m_p=\max_{i\leq p}\sum_{j\leq p}|\sigma_{u,ij}|^q, \quad m_p=o(p),\quad\text{for some}\quad q\in[0,1).
\end{equation}
Under (\ref{eq12}), the test statistics $\by^{\star}=\sqrt n\overline\by$ follow the approximate factor model
\begin{equation} \label{eq13}
\by^{\star}=\bmu^{\star}+\bB\bff^{\star}+\bu^{\star} \sim N(\bmu^{\star}, \bSigma),
\end{equation}
where $\bmu^{\star}=\sqrt{n}\bmu$, $\bff^{\star}=\sqrt{n}\bar\bff$ and $\bu^{\star}=\sqrt{n}\bar\bu$ with
$\bar\bff$ and $\bar\bu$ being the corresponding mean vector.

Fan, Liao \& Mincheva (2013) developed a method called POET to estimate the unknown $\bSigma$ based on samples $\{\by_i\}_{i=1}^n$ in (\ref{eq12}). The basic idea is to take advantage of the factor model structure and the sparsity of the covariance matrix of idiosyncratic noises.  Their idea combined with PFA in Fan, Han \& Gu (2012) yields the following {\bf POET-PFA method}.
\begin{enumerate}
 \item Compute sample covariance matrix $\widehat\bSigma$ and decompose $\widehat{\bSigma}=\sum_{i=1}^p\tilde{\lambda}_i\tilde{\bgamma}_i\tilde{\bgamma}_i^T$, where $\{\tilde{\lambda}_i\}$ and $\{\tilde{\bgamma}_i\}$ are the eigenvalues and eigenvectors of $\widehat{\bSigma}$.  Apply a thresholding method to $\sum_{i=k+1}^p\widetilde{\lambda}_i\widetilde{\bgamma}_i\widetilde{\bgamma}_i^T$ to obtain $\widehat{\bSigma}_u^{\mathcal{T}}$ (e.g. the adaptive thresholding method in Supplementary Materials). Set $\widehat{\bSigma}_{\text{POET}}=\sum_{i=1}^k
     \widetilde{\lambda}_i\widetilde{\bgamma}_i\widetilde{\bgamma}_i^T
     +\widehat{\bSigma}_u^{\mathcal{T}}$.

 \item Apply singular value decomposition to $\widehat{\bSigma}_{\text{POET}}$. Obtain its eigenvalues
       $\widehat\lambda_1,\cdots,\widehat\lambda_K$ in non-increasing order and the associated
       eigenvectors $\widehat\bgamma_1,\cdots,\widehat\bgamma_K$.

 \item Construct $\widehat\bB=(\widehat \lambda_1^{1/2}\widehat\bgamma_1,
       \cdots, \widehat\lambda_K ^{1/2}\widehat\bgamma_K)$ and compute the least-squares
       $\widehat\bff^{\star} = (\widehat\bB^T \widehat\bB)^{-1} \widehat\bB^T \sqrt{n} \overline{\bX}$, which is the least-squares estimate from (\ref{eq13}) with $\bmu^\star$ ignored.

 \item With $\widehat \bb_i^T$ denoting the $i^{th}$ row of $\widehat \bB$, compute
 \begin{equation}\label{h2}
       \widehat{\text{FDP}}_{\text{POET}}(t)=\sum_{i=1}^p[\Phi(\widehat a_i(z_{t/2}
       +\widehat\bb_i^T\widehat\bff^{\star}))+\Phi(\widehat a_i(z_{t/2}-\widehat\bb_i^T\widehat\bff^{\star}))]/R(t)
 \end{equation}
       for some threshold value $t$, where $\widehat a_i=(1-\|\widehat\bb_i\|^2)^{-1/2}$.
\end{enumerate}


The convergence rate of $\widehat{\textrm{\FDP}}_{\text{POET}}(t)$ is as follows. Note that under Assumptions 1-4 in Supplementary Materials, Lemma 2 there holds with high probability.  Let us call this event ${\cal E}^*$. Let ${\cal E}_1$ be the event that condition C1 and C4 are satisfied. 
\begin{theorem}
For POET-PFA method,  we have
\begin{eqnarray*}
\Big|\widehat{\textrm{\FDP}}_{\text{POET}}(t)-\textrm{\FDP}_A(t)\Big|&=&O_p\Big(p^{\theta}(k(\omega_p + m_p\omega_p^{1-q}p^{-1})+\|\bmu^{\star}\|p^{-1/2}\big) \Big),
\end{eqnarray*}
on the event ${\cal E}_1 \cap {\cal E}^*$,
where $\omega_p=p^{-1/2}+\sqrt{\log p/n}$.
\end{theorem}

Theorem 4 can be considered as a corollary of Theorems 1 and 3. However, since POET-PFA is the method that we recommend, we would like to state it as a theorem to emphasize its importance. It is worth noting that here $|\FDP_{oracle}(t)-\FDP(t)|=O_p(p^{\theta}m_p^{1/2}p^{-1/2})$ by the examination of the proof of Proposition 2 in Fan, Han \& Gu (2012). 

\subsection{Dependence-Adjusted Procedure}
The p-value of each test is determined completely by individual $Z_i$, which ignores the correlation structure.  This method can be inefficient, as Fan, Han \& Gu (2012) pointed out.  This section shows how to use dependent structure to improve the power of the test and how to provide an alternative ranking of statistical significance from ranking of $\{|Z_i|\}_{i=1}^p$ under dependence.

Under model (\ref{eq5}),  $a_i(Z_i-\bb_i^T\bW)\sim N(a_i\mu_i,1)$. Since $a_i>1$, this increases the strength of signals and provides an alternative ranking of the significance of each hypothesis.
Indeed, the P-value based on this adjusted test statistic is now
$2\Phi(-|a_i(Z_i-\bb_i^T\bW)|)$ and the null hypothesis $H_{i0}$ is rejected when it is no larger than $t$.   In other words, the  critical region is $|a_i(Z_i-\bb_i^T\bW)| \leq  |z_{t/2}|$. When the covariance matrix $\bSigma$ is unknown, we calculate the p-values as $P_i=2\Phi(-|\widehat{a}_i(Z_i-\widehat{\bb}_i^T\widehat{\bW}|)$, where $\widehat{a}_i$, $\widehat{\bb}_i$ and $\widehat{\bW}$ have been defined in (\ref{eq10}) and (\ref{eq11}). The theoretical investigation of this procedure is beyond the scope of the current paper. We will show in simulation studies that this dependence-adjusted procedure is still more powerful than the fixed threshold procedure.
\section{Simulation Studies}
In the simulation studies, we consider the dimensionality $p=1000$, the sample size $n=50, 100, 200$,  the number of false nulls $p_1=50$, the threshold value $t=0.01$ and the number of simulation round 500, unless stated otherwise. The data are generated from $\bx_i\sim N_p(\bmu, \bSigma)$ except in the following model 3. The signal strength $\mu_i=1$ for $i=1, \cdots, 50$ and 0 otherwise. To investigate the effect of signal strength, we also consider nonzero $\mu_i$ as 0.8 and 1.2. To save space, these results are shown in the supplementary materials. We estimate the unknown number of factors $k$ for POET-PFA by the data-driven eigenvalue ratio method described in Section 3.1 with $k_{\max}=\lfloor 0.2n\rfloor$. To demonstrate the wide applicability of POET-PFA compared with other methods, we consider 8 different model settings for dependence structures in Table 1 as well as Tables 3 \& 4 in the supplementary materials:

\vspace{0.06in}
\noindent \textbf{Model 1: Strict Factor Model}. Consider a 3-factor model
$$
  \bx_i=\bmu+\bB\bff_i+\bu_i, \quad \bff_i\sim N_3(0,\bI_3) \quad \mbox{ indep. of } \quad \bu_i\sim N_p(0,\bSigma_u),
$$
Each entry of the factor loading matrix $\bB_{ij}$ is an independent realization from the uniform distribution $U(-1,1)$. In addition, $\bSigma_u=\bI_p$.

\vspace{0.06in}
\noindent\textbf{Model 2: Approximate Factor Model}. The model set up is the same as Model 1, except that we construct $\bSigma_u$ as follows.  First apply the method in Fan, et al. (2013) to create a covariance matrix $\bSigma_1$, which was calibrated to the returns of S\&P500 constituent stocks.  We omit the details. Then we construct a symmetric banded matrix $\bSigma_2$. For the $(i,j)$th element, if $i\neq j$ and $|i-j|\leq25$, set the element as 0.4 and zero otherwise. Next we construct a symmetric matrix $\bSigma_3$ as the nearest positive definite matrix of $\bSigma_1+\bSigma_2$ by the algorithm of Higham (1988). Finally the covariance matrix $\bSigma_u$ is set as $0.5\bSigma_3$.

\vspace{0.06in}
\noindent \textbf{Model 3: Non-Normal Model}. Consider a 5-factor model $\bx_i=\bmu+\bB\bff_i+\bu_i$. $\bB$ is generated similarly to Model 1, but each element of $\bff_i$ and each element of $\bu_i$ are independent realizations from $\sqrt{2/3}t_6$ where $t_6$ is a $t$ distribution with degrees of freedom as 6. Model 3 is constructed to show the performance of POET-PFA even when the normality assumption for the data-generating process is violated.

\vspace{0.06in}
\noindent\textbf{Model 4: Cluster Model}. We first generate a $p-$dimensional vector $\Lambda$, where the first 4 elements are independent realizations from the uniform distribution $U(160, 190)$, the next 10 elements are independently from $U(8, 12)$ and the rest are independently from $U(0.1, 0.3)$. Next we generate a $p\times p$ matrix $\bQ$ in which each element is an independent realization from $N(0,1)$. Let $\bGamma$ be the matrix, consisting of eigenvectors of $\bQ\Lambda\bQ^T$. Finally,  let $\bSigma=\bGamma\Lambda\bGamma^T$. Model 4 is designed against the eigengap condition in Theorem 3 and also test the robustness of determining number of factors.

\vspace{0.06in}
\noindent \textbf{Model 5: Long Memory Autocovariance Model}. Consider $\bSigma$ where each element is defined as $\Sigma_{ij}=0.5*[||i-j|+1|^{2H}-2|i-j|^{2H}+||i-j|-1|^{2H}], 1\leq i, j\leq p$ with $H=0.9$. Model 5 is from Bickel \& Levina (2008a) and has also been recently considered by Huang \& Fryzlewicz (2015) for strong long memory dependence.

\vspace{0.06in}
\noindent\textbf{Model 6: Normal Perturbation Model}. Consider a symmetric matrix $\bQ$ with diagonal elements as 1 and each off-diagonal element as independent realization from $N(0.5, 0.1)$. Let $\bSigma$ be the nearest positive definite matrix of $\bQ$ based on the algorithm in Higham (1988). Model 6 is constructed lacking an apparent factor model pattern.

\vspace{0.06in}
\noindent \textbf{Model 7: Sparse Precision Matrix Model I}. Consider the precision matrix $\bOmega=\diag(\bA_1,\bA_2)$, where $\bA_2=4\bI_{p/2\times p/2}$, $\bA_1=\bB+\epsilon\bI_{p/2\times p/2}$. $\bB$ is a symmetric matrix where each element $b_{ij}$ takes value 0.5 with probability 0.1 and takes value 0 with probability 0.9. $\epsilon=\max(-\lambda_{\min}(\bB), 0)+0.01$ to ensure that $\bA_1$ is positive definite. Finally, let $\bSigma=(\bOmega)^{-1}$. Construction of $\bA_1$ is from Rothman, et al (2008) for a sparse precision matrix structure. 

\vspace{0.06in}
\noindent\textbf{Model 8: Sparse Precision Matrix Model II}. Consider the precision matrix $\bOmega=\diag(\bA_1,\bA_2)$ similarly to Model 7 except that each $b_{ij}$ takes value uniformly in $[0.3, 0.8]$ with probability 0.2 and takes value 0 with probability 0.8. Finally, let $\bSigma=(\bOmega)^{-1}$. The sparsity structure in Model 8 is from Cai \& Liu (2011) but we consider this sparsity structure for the precision matrix. The final $\bSigma$ is quite different from Model 7.

\textbf{Comparison with other methods for estimating FDP}.

\begin{table}
\caption{\label{Tab2} Empirical mean absolute error between true $\FDP(t)$ and $\widehat{\FDP}(t)$. The nonzero $\mu_i=1$. The results are in percent. 
}
\centering

\begin{tabular}{cccccccc}
\hline\hline
       & POET-PFA & Efron  & FAMT  & FAMT-PFA  & HF-PFA & SS-PFA  & LW-PFA  \\
\hline\hline
Model 1       &           &          &            &         &          &          &     \\
     $n=50$   & 4.39   & 19.72    & 11.48    & 5.90  & 5.40   & 6.95  & 5.94   \\
     $n=100$  &  3.66   & 19.53    & 10.26     & 4.91  &  4.83   & 4.90    & 4.56   \\
     $n=200$  &  3.34   &  19.58    & 11.86   &  5.33  & 3.60   & 3.85    & 3.71  \\
\hline
Model 2    &   &   &   &  &     &    &   \\
     $n=50$  & 5.09  & 17.49  & 10.15  & 5.56  & 5.69   & 7.49   & 6.93 \\
     $n=100$ & 3.93   & 17.80    & 11.42   & 5.61  & 5.53    & 5.28    & 5.14 \\
     $n=200$ & 3.81   & 18.37   & 10.49  & 5.17    & 5.11    & 4.24   & 4.20 \\
\hline
Model 3  &      &     &       &    &     &     &      \\
     $n=50$     & 5.61    & 15.05    & 12.23    & 6.67    & 6.29     & 7.57     & 6.50  \\
     $n=100$   & 4.24    & 14.37    & 12.69    & 6.29      &5.22      & 5.87    & 5.35  \\
     $n=200$   & 3.84      & 14.63     & 12.27    & 5.54    & 4.55   &  4.77  &  4.60  \\
\hline
Model 4  &      &     &       &    &    &    &    \\
     $n=50$     & 4.62      & 19.49     & 11.40      & 6.62  &  5.50  & 7.26    & 7.10    \\
     $n=100$   & 4.07      & 19.01       & 11.25       & 6.75   & 5.41   & 4.97   & 5.09    \\
     $n=200$   & 3.48      & 18.71      & 10.14       &6.05   & 3.80   & 3.94    & 3.98     \\
\hline
Model 5  &      &     &       &    &     &     &     \\
     $n=50$     &  5.44     & 10.46    & 10.09   & 5.18    & 6.95    & 7.38  & 5.66  \\
     $n=100$   & 5.65   & 10.57   & 10.64   & 5.33  & 6.81  & 6.46     & 5.86   \\
     $n=200$   & 5.29  & 10.64   & 10.76   & 4.65    & 7.03   & 5.78    & 5.47   \\
\hline
Model 6  &      &     &       &     &     &     &      \\
     $n=50$     & 4.60    & 10.12   & 9.84   & 4.83    & 4.73    & 6.08    & 4.60  \\
     $n=100$   & 4.03    & 9.44   & 8.59    & 3.89    & 3.67    & 4.82    & 4.03  \\
     $n=200$   & 4.13   & 9.36   & 10.20    & 4.40    & 4.83    & 4.45   & 4.13   \\
\hline
Model 7  &      &     &       &    &    &    &       \\
     $n=50$     &4.50   & 10.18   & 5.88     & 4.68    & 4.98    & 6.24   & 4.63 \\
     $n=100$   & 4.30   & 10.33   &6.19      & 4.77   & 4.66    & 5.29   & 4.43  \\
     $n=200$   & 4.13   & 9.99   & 6.17     & 4.58    & 5.21   & 4.66     & 4.21   \\
\hline
Model 8  &      &     &       &      &     &     &    \\
     $n=50$     & 4.53   & 11.66    & 6.35    & 4.77    & 5.76    &  6.72   & 5.02   \\
     $n=100$   & 4.25   & 11.13   & 6.30    & 4.81     & 5.16    & 5.26    & 4.41   \\
     $n=200$   & 4.02   & 10.62   & 6.01    & 4.42     & 6.07    & 4.59     & 4.14   \\
\hline\hline
\end{tabular}
\end{table}
We compare our POET-PFA method with the methods in Efron (2007) and Friguet, Kloareg \& Causeur (2009). The latter assumes a strict factor model and uses the expectation-maximization (EM) algorithm to estimate the factor loadings $\bB$ and the common factors $\{\bff_i\}_{i=1}^n$. Correspondingly, they constructed an estimator for $\FDP(t)$ based on their factor model and multiple testing (FAMT) method.   To see how well the EM-algorithm estimates factor loadings $\widehat{\bB}$, we include FAMT-PFA, which replaces $\widehat \bB$ in step 4 of our POET-PFA method with that computed by the EM algorithm, for comparison. In the above simulations, we used the R package ``FAMT" from Friguet, Kloareg \& Causuer (2009) to obtain the EM based estimators $\widehat{\bB}$ and $\{\widehat{\bff}\}_{i=1}^n$. We further consider other methods for estimating the unknown $\bSigma$ rather than POET and compare the performance of corresponding $\widehat{\FDP}$. Exploration in this direction could be endless, and we only consider three representative types of shrinkage estimators here: Huang \& Fryzlewicz(2015) (HF), Schafer \& Strimmer (2005) (SS) and Ledoit \& Wolf (2003) (LW). Note that although these three methods do not involve estimating the number of factors $k$ for the covariance matrix step, they still need to estimate $k$ for the PFA step. Therefore, we apply the eigenvalue ratio method to their methods for a fair comparison with our POET-PFA. The results in HF-PFA are based on 50 simulation rounds by its cross-validation based algorithm ``NOVELIST". Other results are still based on 500 simulation rounds.

In Table 1, we calculate the empirical mean absolute error (the absolute difference between the true FDP and $\widehat{\FDP}$) for the seven methods. We recall that the FDP is a quantity measured in percent and therefore the measurement unit for the mean absolute error reported in Table 1 is percent. Generally, when the sample size increases, the mean absolute error of POET-PFA tends to be smaller. The results in Model 6 seems to be a violation of this statement. However, considering that $\FDP_A(t)$ tends to be an upper bound of FDP, the results here are still reasonable. Overall, our POET-PFA method performs the best compared with other six methods, in terms of producing smaller mean absolute error. In Model 5, FAMT-PFA outperforms POET-PFA, however, further investigation shows that the average of $\widehat{\FDP}$ by FAMT-PFA is an underestimate of the true FDR, while our POET-PFA provides an overestimate, which is better for practical FDR control.  Results for signal strength as 0.8 and 1.2 are shown in Tables 3 \& 4 in the supplementary materials, and are consistent with the findings in Table 1 here.

Figure 1 further demonstrates the performance of our POET-PFA method involving least squares estimation compared with Efron's method,  FAMT, and FAMT-PFA under Models 1 \& 2.  The sample size $n=50$. Our POET-PFA method approximates the true $\FDP(t)$ well. Efron's method captures the general trend of $\FDP(t)$ when the true values are relatively small and deviates away from the true values in the opposite direction when $\FDP(t)$ becomes large. FAMT-PFA performs much better than FAMT, but still could not capture the true value when $\FDP(t)$ is large. Comparison under Models 3-8 are shown in the supplementary materials.
\begin{figure}[h!!!]
\begin{center}
\scalebox{0.56}{\includegraphics{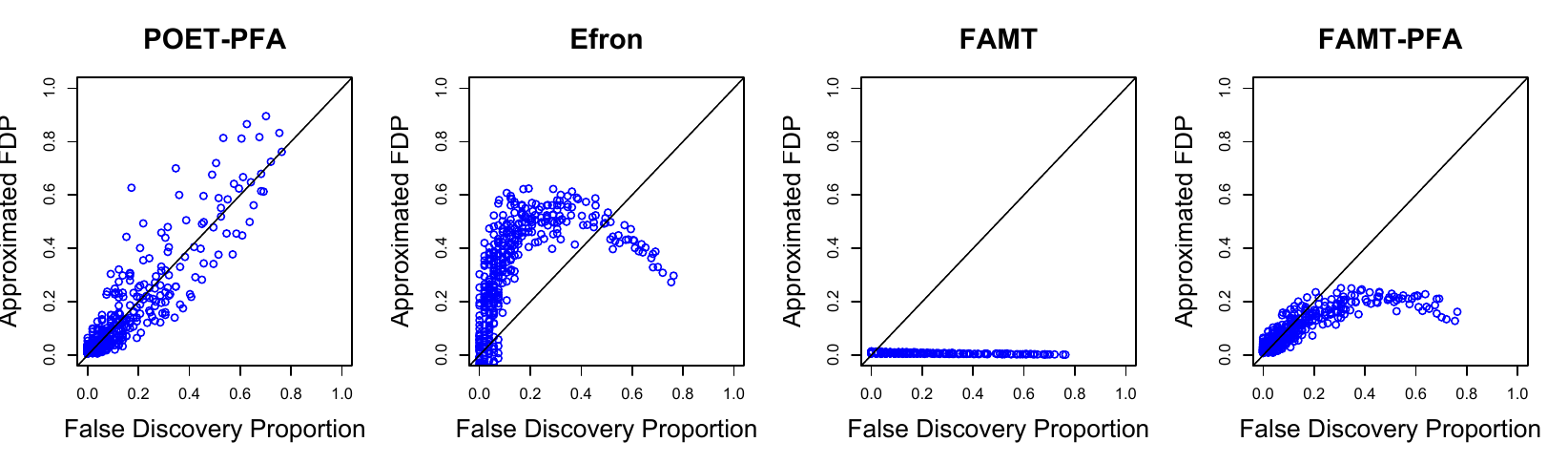}}
\scalebox{0.56}{\includegraphics{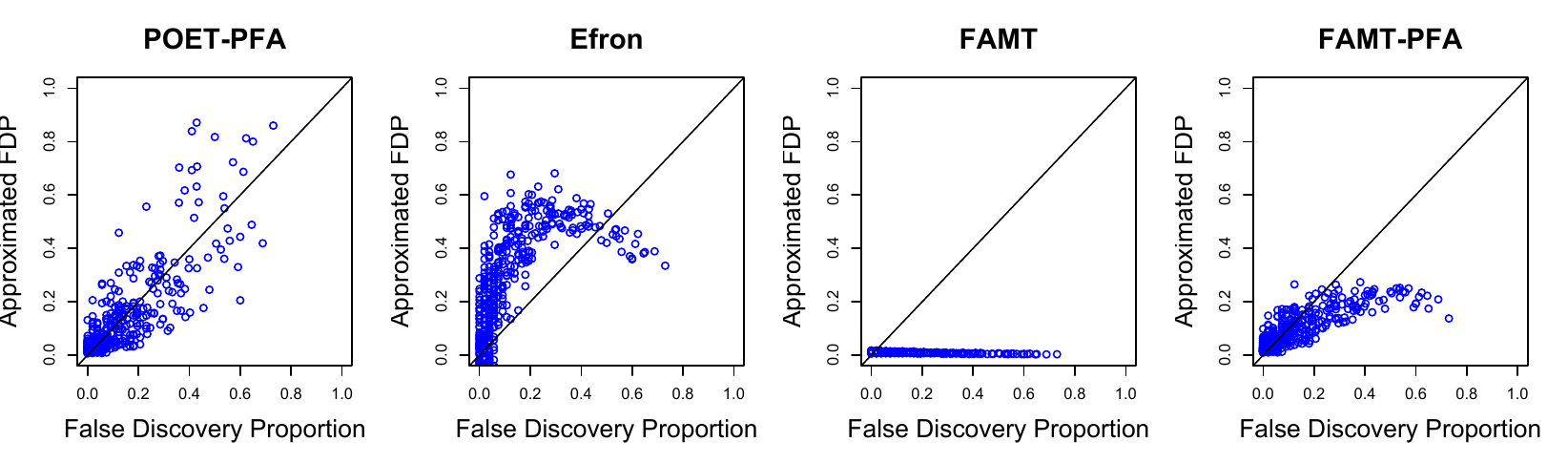}}
\end{center}
\vspace{-0.3cm}
\caption{Comparison of realized values of False Discovery Proportion with $\widehat{\FDP}(t)$. Top panel corresponds to Model 1 and bottom panel corresponds to Model 2. $n=50$.  }
\end{figure}

\textbf{Dependence adjusted testing procedure}.
We compare the dependence-adjusted procedure described in section 3.5 with the fixed threshold procedure, that is, compare the $|Z_i|$ with a universal threshold without using the correlation information. Define the false negative rate $\FNR=E[T/(p-R)]$ where $T$ is the number of falsely accepted null hypotheses. With the same FDR level, a procedure with smaller false negative rate is more powerful. Since the advantage of dependence-adjusted procedure can be better demonstrated by an apparent factor-model structure, the following Table \ref{e5} only considers Models 1 \& 2. In Table \ref{e5},  we fix threshold value $t=0.001$ and reject the hypotheses when the dependence-adjusted $p$-values is smaller than 0.001. Then we find the corresponding threshold value for the fixed threshold procedure such that the FDR in the two testing procedures are approximately the same. To highlight the advantage of dependence-adjusted procedure, we reset $\bSigma_u$ as $0.1\bSigma_3$. The FNR for the dependence-adjusted procedure is smaller than that of the fixed threshold procedure, which suggests that dependence-adjusted procedure is more powerful. In Fan, Han \& Gu (2012), they have shown numerically that if the covariance is known, the advantage of dependence-adjusted procedure is even more substantial. Note that in Table \ref{e5}, $p_1=200$ compared with $p=1000$, implying that the better performance of the dependence-adjusted procedure is not limited to sparse situation. This is expected since subtracting common factors out make the problem have a higher signal to noise ratio.

\begin{table}
\caption{\label{e5}Comparison of Dependence-Adjusted Procedure with Fixed Threshold Procedure under approximate factor model and strict factor model.  The nonzero $\mu_i$ are simulated from $U(0.1,0.5)$ and $p_1=200$.  }
\begin{tabular}{ccccccc}
  \hline\hline
         &\multicolumn{3}{c}{Fixed Threshold Procedure} &\multicolumn{3}{c}{Dependence-Adjusted Procedure}\\
  \cline{2-4} \cline{5-7}
         &FDR   &FNR   &Threshold  & FDR  &FNR  &Threshold\\
   \hline
   Model 1              &    &   &     &    &    &  \\
   $n=50$          &3.21\%   &14.54\%    &0.0026     &3.24\%    &1.96\%    &0.001\\
   $n=100$        &2.48\%   &9.53\%    &0.0048   &2.46\%    &0.54\%    &0.001\\
   $n=200$        &2.85\%     &4.65\%      &0.0074      &2.89\%       &0.08\%       &0.001\\
  \hline
  Model 2  &   &   &    &    &    &   \\
   $n=50$	       &2.64\%    &15.03\%    &0.0028    &2.66\%    &2.40\%    &0.001     \\
   $n=100$       &1.86\%   &10.56\%    &0.0034    &1.85\%    &0.70\%   &0.001 \\
   $n=200$       &1.86\%    &5.65\%    &0.0044     &1.86\%     &0.09\%    &0.001 \\
\hline
\end{tabular}
\end{table}

Additional simulation results regarding comparison with known covariance matrix case 
can be found in Supplementary Materials.  The basic findings are that under apparent factor model structure the estimation errors of covariance matrix have limited impact (see Figures 1 \& 2 there) and methods [the least-absolute deviation (\ref{eq10}), the least-squares estimate (\ref{eq11}), SCAD (\ref{eq8})] for extracting unobservable realized latent factors are all effective. 

\section{Data Analysis}
In a well-known breast cancer study (Hedenfalk et al., 2001, Efron, 2007), scientists compared gene expression levels in 15 patients. These observed gene expression levels have one of the two different genetic mutations, BRCA1 and BRCA2, known to increase the lifetime risk of hereditary breast cancer. The study included 7 women with BRCA1 and 8 women with BRCA2. Let $\bX_1,\cdots,\bX_n$, $n=7$ denote the microarray of expression levels on the $p=3226$ genes for the first group, and $\bY_1,\cdots,\bY_m$, $m=8$ for that of the second group, so each $\bX_i$ and $\bY_i$ are $p$-dimensional column vectors. Understanding the groups of genes that are expressed significantly differently in breast cancers can help scientists identify cases of hereditary breast cancer on the basis of gene-expression profiles.

Assume the gene expressions of the two groups on each microarray are from two multivariate normal distributions with (potentially) different mean vector but the same covariance matrix, namely, $\bX_i \sim N_p(\bmu^X,\bSigma)$ for $i=1,\cdots,n$ and $\bY_i \sim N_p(\bmu^Y,\bSigma)$ for $i=1,\cdots,m$. Then identifying differentially expressed genes is essentially a multiple hypothesis test on $H_{0j}:\mu^X_j=\mu^Y_j$ vs $H_{1j}:\mu^X_j\neq \mu^Y_j$, $j=1,\cdots,p$. Consider the test statistics $\bZ=\sqrt{nm/(n+m)}(\overline\bX-\overline\bY)$ where $\overline\bX$ and $\overline\bY$ are the sample averages. Then we have $\bZ\sim N_p(\bmu,\bSigma)$ with $\bmu=\sqrt{nm/(n+m)}(\bmu^X-\bmu^Y)$, and the above two-sample comparison problem is equivalent to simultaneously testing $H_{0j}:\mu_j=0$ vs $H_{1j}:\mu_j\neq0$, $j=1,\cdots,p$ based on $\bZ$ and the unknown covariance matrix $\bSigma$. It is also reasonable to assume that a large proportion of the genes are not differentially expressed, so that $\bmu$ is sparse.

Factor model structure has gained increasing popularity among biologists in the past decade, since it has been widely acknowledged that gene activities are usually driven by a small number of latent variables. See, for example, Friguet, Kloareg \& Causeur (2009) and  Desai \& Storey (2012) for more details.  We therefore apply the POET-PFA procedure (see Section 3.4) to the dataset to obtain $\widehat{\text{FDP}}_{\text{POET}}(t)$ for a given threshold value $t$. We apply the eigenvalue ratio method as in Section 3.1 to estimate the unknown number of factors. The estimated $k$ is 1 based on the sample data. Due to the small sample size, this estimate could deviate away from the true value. Therefore, we also report the results for $k=2,3,4,5$. The results of our analysis are depicted in Figure 2. As can be seen, both $\widehat{\text{FDP}}_{\text{POET}}(t)$ and $\widehat V(t)$ increase with larger $R(t)$, and $\widehat{\text{FDP}}_{\text{POET}}(t)$ is fairly close to zero when $R(t)$ is below 200, suggesting that the rejected hypotheses in this range have high accuracy to be the true discoveries. Secondly, even when as many as 1000 hypotheses, corresponding to almost $1/3$ of the total number, have been rejected, the estimated FDPs are around $25\%$. Finally it is worth noting that although our procedure seems robust under different choices of number of factors, the estimated FDP tends to be relatively small with larger number of factors. We also apply the dependence-adjusted procedure to the data. The relationship of $\widehat{\FDP}$ and number of total rejections are summarized in Figure 5 in the supplementary materials. Compared with Figure 2, the $\widehat{\FDP}$ tends to be smaller with the same amount of total rejections. The same phenomenon also happens to the estimated number of false rejections.  This is consistent with the fact that the factor-adjusted test is more powerful.
\begin{figure}[h!!!]
\begin{center}
\scalebox{0.52}{\includegraphics{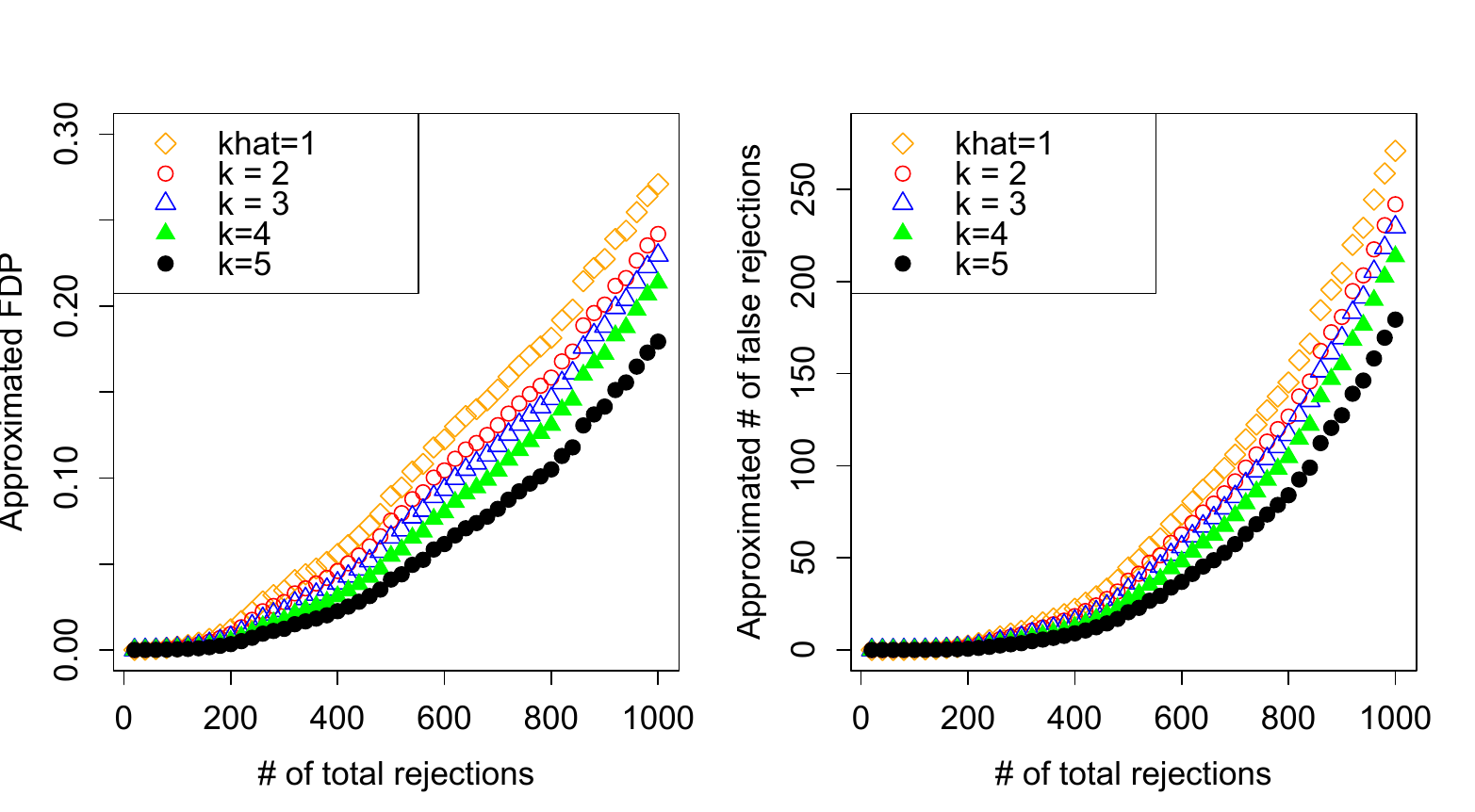}}
\end{center}
\vspace{-0.3cm}
\caption{The approximated false discovery proportion and the approximated number of false discoveries as functions of
the number of total discoveries for $p=3226$ genes, where the estimated number of factors is 1 compared with other choices $k=2,3,4,5$. }\label{Fig100}
\end{figure}
We conclude our analysis by presenting the list of 40 most significantly differentially expressed genes in Table 4 and Table 5 of Supplementary Materials with POET-PFA method and the dependence-adjusted procedure respectively.  Table 5 provides an alternative ranking of statistically significantly expressed genes for biologists, which have a lower false discovery proportion than the conventional method presented in Table 4.

\section{Appendix}

\textbf{Proof of Theorem 1:} First of all, note that by (\ref{eq10}), we have
\begin{equation} \label{eq15}
   \hat{\bB}\widehat{\bW} = \hat{\bB} (  \hat{\bB}^T  \hat{\bB})^{-1}  \hat{\bB}^T \bZ =
   (\sum_{i=1}^k\widehat{\bgamma}_i\widehat{\bgamma}_i^T)\bZ.
\end{equation}
Similarly, let $\bB = (\sqrt{\lambda_1}\bgamma_1,\cdots,\sqrt{\lambda_k}\bgamma_k)$ and
$\widetilde{\bW} = (\bB^T  \bB)^{-1} \bB^T \bZ$.  Then,
\begin{eqnarray} \label{eq16}
   \bB\widetilde{\bW} =
   (\sum_{i=1}^k {\bgamma}_i {\bgamma}_i^T)\bZ.
\end{eqnarray}
Denote by $\FDP_1(t)$ the estimator in equation (\ref{eq7}) with using the infeasible estimator $\widetilde{\bW}$.  Then,
$$
\widehat{ \FDP}_U(t)  - \FDP_A(t) = [\widehat{\FDP}_U(t)  - \FDP_1(t)] + [\FDP_1(t)  - \FDP_A(t)].
$$
We will bound these two terms separately.

Let us deal with the first term.
Define
\begin{eqnarray*}
\Delta_1&=&\sum_{i=1}^p\Big[\Phi(\widehat{a_i}(z_{t/2}+\widehat{\bb}_i^T\hw))-\Phi(a_i(z_{t/2}+\bb_i^T\widetilde{\bw}))\Big]\\
\Delta_2&=&\sum_{i=1}^p\Big[\Phi(\widehat{a_i}(z_{t/2}-\widehat{\bb}_i^T\hw))-\Phi(a_i(z_{t/2}-\bb_i^T\widetilde{\bw}))\Big].
\end{eqnarray*}
Then, we have
\begin{equation}
\widehat{\FDP}_{U}(t)-\FDP_1(t)=(\Delta_1+\Delta_2)/R(t).   \label{eq17}
\end{equation}

We now deal with the term $\Delta_1=\sum_{i=1}^p\Delta_{1i}$, in which
\begin{eqnarray*}
\Delta_{1i}
           &=&\Phi(\widehat{a_i}(z_{t/2}+\widehat{\bb}_i^T\widehat{\bw}))-\Phi(\widehat{a_i}(z_{t/2}+\bb_i^T\widetilde{\bw}))\\
           &&+\Phi(\widehat{a_i}(z_{t/2}+\bb_i^T\widetilde{\bw}))-\Phi(a_i(z_{t/2}+\bb_i^T\widetilde{\bw}))\\
           &\equiv&\Delta_{11i}+\Delta_{12i}.
\end{eqnarray*}
$\Delta_2$ can be dealt with analogously and hence omitted.
For $\Delta_{12i}$, by the mean-value theorem, there exists $a_i^{*}\in(a_i,\widehat{a_i})$ such that $\Delta_{12i}=\phi(a_i^{*}(z_{t/2}+\bb_i^T\widetilde{\bw}))(\widehat{a}_i-a_i)(z_{t/2}+\bb_i^T\widetilde{\bw})$. Since $a_i>1$ and $\widehat{a_i}>1$, we have $a_i^*>1$ and hence $\phi(a_i^{*}(z_{t/2}+\bb_i^T\widetilde{\bw}))|z_{t/2}+\bb_i^T\widetilde{\bw}|$ is bounded.  In other words, $|\sum_{i=1}^p \Delta_{12i}| \leq C \sum_{i=1}^p|\widehat{a}_i-a_i|$, for a generic constant $C$. Using the definition of $\widehat{a}_i$ and $a_i$, we have
\begin{equation*}
|\widehat{a}_i-a_i|=|(1- \|\widehat{\bb}_{i}\|^2)^{-1/2}
- (1- \|\bb_{i}\|^2)^{-1/2}|.
\end{equation*}
Using the mean-value theorem again, together with the assumption (C4), we have
\begin{equation*}
|(1- \|\widehat{\bb}_{i}\|^2)^{-1/2}
- (1- \|\bb_{i}\|^2)^{-1/2}| \leq C ( \|\widehat{\bb}_{i}\|^2 - \|\bb_{i}\|^2 ).
\end{equation*}
Let $\bgamma_h=(\gamma_{1h},\cdots,\gamma_{ph})^T$ and $\widehat{\bgamma}_h=(\widehat{\gamma}_{1h},\cdots,\widehat{\gamma}_{ph})^T$. Then
\begin{eqnarray*}
\sum_{i=1}^p  \Bigl |  \|\widehat{\bb}_{i}\|^2 - \|\bb_{i}\|^2 \Bigr |
&=&\sum_{i=1}^p \Bigl |\sum_{h=1}^k(\widehat{\lambda}_h-\lambda_h)
\widehat{\gamma}_{ih}^2 + \sum_{h=1}^k\lambda_h(\widehat{\gamma}_{ih}^2-\gamma_{ih}^2) \Bigr |\\
&\leq&\sum_{h=1}^k |\widehat{\lambda}_h-\lambda_h| + \sum_{h=1}^k \lambda_h \sum_{i=1}^p|\widehat{\gamma}_{ih}^2-\gamma_{ih}^2|,
\end{eqnarray*}
where we used $\sum_{i=1}^p \hat{\gamma}_{ih}^2 = 1$.
The second term of the last expression can be bounded as
\begin{eqnarray*}
\sum_{i=1}^p|\widehat{\gamma}_{ih}^2-\gamma_{ih}^2|&\leq&
\Bigl ( \sum_{i=1}^p|\widehat{\gamma}_{ih}-\gamma_{ih}|^2
\sum_{i=1}^p|\widehat{\gamma}_{ih}+\gamma_{ih}|^2 \Bigr )^{1/2}\\
    &\leq& \| \widehat{\bgamma}_h-\bgamma_h \| \bigl \{ 2\sum_{i=1}^p (\widehat{\gamma}_{ih}^2+\gamma_{ih}^2)\bigr \}^{1/2}\\
    &=&2 \|\widehat{\bgamma}_h-\bgamma_h \|.
\end{eqnarray*}
Combining all the results that we have obtained, we have concluded that
\begin{equation} \label{eq18}
  |\sum_{i=1}^p \Delta_{12i}| \leq C \Bigl (\sum_{h=1}^k |\widehat{\lambda}_h-\lambda_h| + \lambda_h  \|\widehat{\bgamma}_h-\bgamma_h \| \Bigr ).
\end{equation}
Therefore, by using  $\sum_{h=1}^k\lambda_h<p$ and Assumptions (C2) and (C3), on the event $\cal{E}$, we conclude that $|\sum_{i=1}^p\triangle_{12i}|=O(p^{1-\min(\nu,\kappa)})$.

We now deal with the term $\Delta_{11i}$.  By the mean-value theorem, there exists $\xi_i$ between $\widehat{\bb}_i^T\widehat{\bw}$ and $\bb_i^T\widetilde{\bw}$ such that $\Delta_{11i}=\phi(\widehat{a_i}(z_{t/2}+\xi_i))\widehat{a_i}(\widehat{\bb}_i^T\widehat{\bw}-\bb_i^T\widetilde{\bw})$. By (C4), $\widehat{a_i}$ is bounded and so is $\phi(\widehat{a_i}(z_{t/2}+\xi_i))\widehat{a}_i$.  Let $\bone$ be a $p$-dimensional vector with each element being 1.  Then, by (\ref{eq15}) and (\ref{eq16}), we have
\begin{equation}
\sum_{i=1}^p | \widehat{\bb}_i^T\widehat{\bw}-\bb_i^T\widetilde{\bw} | \leq \bone^T |\hat{\bB} \widehat{\bw}- {\bB}\widetilde{\bw}|=\bone^T \Big |\sum_{h=1}^k[\widehat{\bgamma}_h \widehat{\bgamma}_h^T -\bgamma_h\bgamma_h^T]\bZ \Big | \leq \sqrt{p} \Big \| \sum_{h=1}^k[\widehat{\bgamma}_h \widehat{\bgamma}_h^T -\bgamma_h\bgamma_h^T] \Big \| \|\bZ\| \label{eq19}
\end{equation}
where $|\ba| = (|a_1|, \cdots, |a_p|)^T$ for any vector $\ba$ and the last inequality is obtained by the Cauchy-Schwartz inequality.

We now deal with the two factors in (\ref{eq19}).  The first factor is easily bounded by
$$
    \sum_{h=1}^k \| \widehat{\bgamma}_h (\widehat{\bgamma}_h - {\bgamma}_h)^T +
        (\widehat{\bgamma}_h - \bgamma_h) \bgamma_h^T \| \leq 2 \sum_{h=1}^k \| \widehat{\bgamma}_h - \bgamma_h\|.
$$
Let $\{\varepsilon_i\}_{i=1}^p$ be a sequence of i.i.d. $N(0, 1)$ random variables.  Then, stochastically, we have
$$
E  \| \bZ \|^2 \leq 2\|\bmu^{\star}\|^2 +2 \sum_{i=1}^p \lambda_i E\varepsilon_i^2.
$$
Therefore, $\|\bZ\|=O_p(\|\mu^{\star}\|+p^{1/2})$.

Substituting these two terms into (\ref{eq19}), we have
$$
   \sum_{i=1}^p | \widehat{\bb}_i^T\widehat{\bw}-\bb_i^T\widetilde{\bw} |
   = O_p\Big (k p^{1/2 - \kappa} (\|\bmu^{\star}\| + p^{1/2}) \Big ).
$$

Therefore,  we can conclude that
\begin{equation}\label{h5}
|\sum_{i=1}^p\Delta_{11i}|= O_p\Big (k p^{1/2 - \kappa} (\|\bmu^{\star}\| + p^{1/2}) \Big ).
\end{equation}
Combination of the results in (\ref{eq18}) and (\ref{h5}) leads to
\begin{equation*}
\Delta_1=O_p(p^{1-\min(\nu, \kappa)})+O_p(kp^{1-\kappa})+O_p(k\|\bmu^{\star}\|p^{1/2-\kappa}).
\end{equation*}

In $\FDP_1(t)$, the least-squares estimator is
\begin{equation}\label{h6}
\widetilde{\bW}=(\bB^T\bB)^{-1}\bB^T\bmu^{\star}+\bW+(\bB^T\bB)^{-1}\bB^T\bK=\bW+(\bB^T\bB)^{-1}\bB^T\bmu^{\star}
\end{equation}
in which we utilize the orthogonality between $\bB$ and $\var(\bK)$. With a similar argument as above, we can show that
\begin{equation*}
\big|\FDP_1(t)-\FDP_A(t)\big|=O\Big(\big|\bone^T\bB(\widetilde{\bW}-\bW)\big|/R(t)\Big),
\end{equation*}
and we have
\begin{equation*}
\Big|(1,\cdots,1)\bB(\widetilde{\bW}-\bW)\Big|=\Big| \bone^T (\sum_{h=1}^k\bgamma_h\bgamma_h^T)\bmu^{\star}\Big|\leq p^{1/2}\|\bmu^{\star}\|\|\sum_{h=1}^k\bgamma_h\bgamma_h^T\|=p^{1/2}\|\bmu^{\star}\|.
\end{equation*}
The proof is now complete.

\textbf{Proof of Theorem 2:} The proof is relegated to the supplementary material due to the space limit.

\textbf{Proof of Theorem 3:} By the triangular inequality,
\begin{equation*}
|\lambda_i-\widehat{\lambda}_{i+1}|\geq\big||\lambda_i-\lambda_{i+1}|-|\lambda_{i+1}-\widehat{\lambda}_{i+1}|\big|
\end{equation*}
By Weyl's Theorem in Lemma 1, $|\lambda_{i+1}-\widehat{\lambda}_{i+1}|\leq\|\widehat{\bSigma}-\bSigma\|$.
Therefore, on the event $\{\|\widehat{\bSigma}-\bSigma\|=O(d_pp^{-\tau})\}$
$$
  |\lambda_i-\widehat{\lambda}_{i+1}|\geq d_p - \|\widehat{\bSigma}-\bSigma\| \geq d_p/2
$$
for sufficiently large $p$. Similarly, we have $|\widehat{\lambda}_{i-1}-\lambda_i|\geq d_p/2$.   By the $\sin \theta$ Theorem in Lemma 1, $\|\bgamma_i-\widehat{\bgamma}_i\|=O(p^{-\tau})$.  Hence, Condition (C2) holds with $\kappa = \tau$.  Using Weyl's Theorem again, we have
$$
\sum_{i=1}^k |\lambda_{i+1}-\widehat{\lambda}_{i+1}|\leq k \|\widehat{\bSigma}-\bSigma\|
= O( k d_p p^{-\tau}).
$$
Hence, (C3) holds with $p^{-\delta} = k p^{-\tau} d_p/p$.
The result now follows from Theorem 1. 

\textbf{Proof of Theorem 4}.  Let $\widetilde{\bB}=(\widetilde{\lambda}_1^{1/2}\widetilde{\bgamma}_1,\cdots,
\widetilde{\lambda}_k^{1/2}\widetilde{\bgamma}_k)$. Note that
\begin{equation} \label{eq24}
\|\widehat{\bSigma}_{\text{POET}}-\bSigma\|\leq
\|\widetilde{\bB}\widetilde{\bB}^T-\bB\bB^T\|+\|\widehat{\bSigma}_u^{\mathcal{T}}-\bSigma_u\|.
\end{equation}
The bound for the second term is given by Lemma 1 of Supplementary Materials.  We now consider the first term in (\ref{eq24}).
By the triangular inequality, it follows that
\begin{eqnarray} \label{eq25}
 \|\widetilde{\bB}\widetilde{\bB}^T-\bB\bB^T\|
&\leq&
\|\bB(\bH^T\bH-\bI_k)\bB^T\|+\|\bB\bH^T(\widetilde{\bB}-\bB\bH^T)^T\|  +\|(\widetilde{\bB}-\bB\bH^T)\bH\bB^T\|
\nonumber \\
    & &
    +\|(\widetilde{\bB}-\bB\bH^T)(\widetilde{\bB}-\bB\bH^T)^T\| \nonumber \\
&\leq& \|\bH^T\bH-\bI_k\|\|\bB\|^2+ 2\|\bB\| \|\bH\| \|\widetilde{\bB}-\bB\bH^T\| +\|(\widetilde{\bB}-\bB\bH^T)\|^2.
\end{eqnarray}

Recall $\{\widetilde{\bb}_j\}_{j=1}^k$ are columns of $\bB$.  Without loss of generality, assume $\{\|\widetilde{\bb}_j\|\}$ are in non-increasing order. Since $\bB^T\bB$ is diagonal, $\bB\bB^T$ has nonvanishing eigenvalues $\{ \|\widetilde{\bb}_j\|^2\}_{j=1}^K$ and $\|\bB\|=\|\widetilde{\bb}_1\|$.  Furthermore, by
Weyl's Theorem in Lemma 1, $\Big|\lambda_i-\|\widetilde{\bb}_i\|^2\Big|\leq\|\bSigma-\bB\bB^T\|=\|\bSigma_u\|$. Since the operator norm is bounded by the $L_1$-norm, we have
\begin{equation}\label{eq26}
\|\bSigma_u\|\leq\max_{i\leq p}\sum_{j=1}^p|\sigma_{u,ij}|^q|\sigma_{u,ii}\sigma_{u,jj}|^{(1-q)/2}\leq m_p.
\end{equation}
Hence, $\|\widetilde{\bb}_i\|^2\leq\lambda_1+m_p=O(p)$.

We are now bounding each term in (\ref{eq25}).  Since the operator norm is bounded by the Frobenius norm, by Lemma 2 of Supplementary Materials, the first term in (\ref{eq25}) is bounded by
$O_p( p \omega_p)$,  the second term in (\ref{eq25}) is of order $O_p(\omega_p \sqrt{p})$ and the third term in (\ref{eq25}) is $O_p(\omega_p^2 )$.
Combination of these results leads to $\|\widetilde{\bB}\widetilde{\bB}^T-\bB\bB\|=O_p(p\omega_p)$. Substituting this into (\ref{eq24}), we have
\begin{equation*}
\|\widehat{\bSigma}_{\text{POET}}-\bSigma\|=O_p(p \omega_p + m_p\omega_p^{1-q}).
\end{equation*}
By Weyl's Theorem in Lemma 1, the conclusion for $|\widehat{\lambda}_i-\lambda_i|$ follows.

Assumption 1 of Supplementary Materials and Weyl's theorem imply that $\lambda_i=c_ip + o(p)$ for $i=1,\cdots,k$ and $c_i$'s are distinct. By the triangular inequality, $|\lambda_i-\widehat{\lambda}_{i+1}|\geq\big||\lambda_i-\lambda_{i+1}|-|\lambda_{i+1}-\widehat{\lambda}_{i+1}|\big|$. By Weyl's Theorem, $|\lambda_{i+1}-\widehat{\lambda}_{i+1}|=o_p(p)$. Therefore, for sufficiently large $n$, $|\lambda_i-\widehat{\lambda}_{i+1}|\geq \widetilde{c}_ip$ for some constant $\widetilde{c}_i>0$ with probability tending to 1. By $\sin\theta$ Theorem, $\|\widehat{\bgamma}_i-\bgamma_i\|=O_p( \omega_p +m_p\omega_p^{1-q}p^{-1})$. With direct application of Theorems 1 \& 3, we have
\begin{equation*}
\Big|\widehat{\textrm{\FDP}}_{\text{POET}}(t)-\textrm{\FDP}_A(t)\Big|=O_p\Big(p^{\theta}\big(k( \omega_p + m_p\omega_p^{1-q}p^{-1})+\|\bmu^{\star}\|p^{-1/2}\big)\Big).
\end{equation*}
The proof is now complete.

\vspace{0.1in}
\noindent\textbf{Acknowledgements}\\
This research was partly supported by NIH Grants R01-GM072611-11 and R01GM100474-04 and NSF Grant DMS-1206464.  We would like to thank Dr. Weijie Gu for early assistance on this project. We also want to thank the Joint Editor Professor Piotr Fryzlewicz, the Past Editor Professor Gareth Roberts, the Associate Editors and anonymous referees for many constructive comments which significantly improve the presentation of the paper.

\vspace{0.1in}

\noindent\textbf{Supplementary Materials}\\
The proof of Theorem 2, some related lemmas, additional numeral results, original codes for simulation studies and data analysis can be found in the supplementary materials.

\section{Related Existing Method and Lemmas}
{\bf Adaptive Thresholding Method}.
This method is a modification of the adaptive thresholding method in Cai \& Liu (2011) and has been introduced in Fan, Liao \& Mincheva (2013). In the approximate factor model, define $\widetilde{\bX}=(\bX_1-\overline{\bX},\cdots,\bX_n-\overline{\bX})$, $\widehat{\bF}^T=(\widehat{\bff}_1,\cdots,\widehat{\bff}_n)$, where the columns of $\widehat{\bF}/\sqrt{n}$ are the eigenvectors corresponding to the $k$ largest eigenvalues of $\widetilde{\bX}^T\widetilde{\bX}$. Let $\widetilde{\bB}=(\widetilde{\lambda}_1^{1/2}\widetilde{\bgamma}_1,\cdots,
\widetilde{\lambda}_k^{1/2}\widetilde{\bgamma}_k)$. Compute $\widehat{\bu}_l=(\bX_l-\overline{\bX})-\widetilde{\bB}\widehat{\bff}_l$,
\begin{equation*}
\widehat{\sigma}_{ij}=\frac{1}{n}\sum_{l=1}^n\widehat{u}_{il}\widehat{u}_{jl}, \quad
\mbox{and} \quad  \widehat{\theta}_{ij}^{\;2} =
\frac{1}{n}\sum_{l=1}^n(\widehat{u}_{il}\widehat{u}_{jl}-\widehat{\sigma}_{ij})^2.
\end{equation*}
For the threshold $\tau_{i,j}=C\widehat{\theta}_{ij} \omega_p$ with a large enough $C$, the adaptive thresholding estimation for $\bSigma_u$ is given by $\widehat{\bSigma}_u^{\mathcal{T}}=(s_{ij}(\widehat{\sigma}_{ij}))_{p\times p}$, where $s_{ij}(\cdot)$ is a general thresholding function (Antoniadis and Fan, 2001) satisfying $s_{ij}(z)=0$ when $|z|\leq\tau_{ij}$ and $|s_{ij}(z)-z|\leq\tau_{ij}$.  Well-known thresholding functions include hard-thresholding estimator $s_{ij}(z) = zI(|z| \geq \tau_{ij})$ and soft-thresholding estimator $s_{ij}(z)=\mathrm{sgn}(z)(|z|-\tau_{ij})_+$.

The following Assumptions 1-4 are from Fan, Liao \& Mincheva (2013).  The results were established for the mixing sequence but it is applicable to the i.i.d. data.
\begin{assumption}
$\|p^{-1}\bB^T\bB-\bOmega\|=o(1)$ for some $k\times k$ symmetric positive definite matrix $\bOmega$ such that $\bOmega$ has $k$ distinct eigenvalues and that $\lambda_{\min}(\bOmega)$ and $\lambda_{\max}(\bOmega)$ are bounded away from both zero and infinity.
\end{assumption}

\begin{assumption}
(i) $\{\bu_l,\bff_l\}_{l\geq1}$ is strictly stationary. In addition, $Eu_{il}=Eu_{il}f_{jl}=0$ for all $i\leq p, j\leq k$ and $l\leq n$. \\
(ii) There exist positive constants $c_1$ and $c_2$ such that $\lambda_{\min}(\bSigma_u)>c_1$, $\|\bSigma_u\|_1<c_2$, and \\
$\min_{i,j} \var(u_{il}u_{jl})>c_1$. \\
(iii) There exist positive constants $r_1$,  $r_2$, $b_1$, and $b_2$ such that for any $s>0$, $i\leq p$ and $j\leq k$,
\begin{equation*}
P(|u_{il}|>s)\leq\exp(-(s/b_1)^{r_1}), \quad\quad\quad P(|f_{jl}|>s)\leq\exp(-(s/b_2)^{r_2}).
\end{equation*}
\end{assumption}

We introduce the strong mixing conditions to conduct asymptotic analysis of the least square estimates. Let $\mathcal{F}_{-\infty}^0$ and $\mathcal{F}_{n}^{\infty}$ denote the $\sigma$-algebras generated by $\{(\bff_s,\bu_s): -\infty\leq s\leq 0\}$ and  $\{(\bff_s,\bu_s): n\leq s\leq \infty\}$ respectively. In addition, define the mixing coefficient
\begin{equation*}
\alpha(n)=\sup_{A\in\mathcal{F}_{-\infty}^0, B\in\mathcal{F}_{n}^{\infty}}|P(A)P(B)-P(AB)|.
\end{equation*}
Note that for the independence sequence, $\alpha(n) = 0$.
\begin{assumption}
    There exists   $r_3>0$ such that $3r_1^{-1}+1.5r_2^{-1}+r_3^{-1}>1$, and $C>0$ satisfying $\alpha(n)\leq \exp(-Cn^{r_3})$ for all $n$.
\end{assumption}
\begin{assumption}
Regularity conditions: There exists $M>0$  such that  for all $i\leq p$, $t\leq n$ and $s\leq n$,\\
(i) $\|\bb_{j}\|_{\max}<M$,\\
(ii)  $\bE[p^{-1/2}({\bu}_s'{\bu}_t-\bE{\bu}_s'{\bu}_t)]^4<M$,\\
(iii) $\bE\|p^{-1/2}\sum_{i=1}^p\bb_iu_{it}\|^4<M$.
\end{assumption}
\begin{lemma}\label{lemma2}
{\bf (Fan, Liao \& Mincheva, 2013, Theorem 1)} \\
Let $\gamma^{-1}=3\gamma_1^{-1}+1.5\gamma_2^{-1}+\gamma_3^{-1}+1$. Suppose $\log p=o(n^{\gamma/6})$ and $n=o(p^2)$. Under Assumptions 1-4,
\begin{equation*}
\|\widehat{\bSigma}_u^{\mathcal{T}}-\bSigma_u\|=O_p(\omega_p^{1-q}m_p).
\end{equation*}
\end{lemma}
Define $\bV=\diag(\widehat{\lambda}_1, \cdots, \widehat{\lambda}_k )$. $\widehat{\bF}^T=(\widehat{\bff}_1,\cdots,\widehat{\bff}_n)$, and $\bH=\frac{1}{n}\bV^{-1}\widehat{\bF}^T\bF\bB^T\bB$, where $\widehat{\bF}$ has been defined in Adaptive Thresholding Method.
\begin{lemma} \label{lemma3}
{\bf (Fan, Liao \& Mincheva, 2013, Lemma C.10 and C.12)} With the same conditions in Lemma 1,
\begin{eqnarray*}
\|\bH\|&=&O_p(1)\\
\|\bH^T\bH-\bI_k\|_F&=&O_p(\frac{1}{\sqrt{n}}+\frac{1}{\sqrt{p}})\\
\|\widetilde{\bB}-\bB\bH^T\|_F^2&=&O_p(\omega_p^2p).
\end{eqnarray*}
\end{lemma}

\begin{lemma}[Fujikoshi \& Mukaihata (1993)]
let $F_n(\cdot)$ and $f_n(\cdot)$ be respectively the cumulative probability function and probability density function of Student's t distribution with $n$ degrees of freedom. Let $\Phi(\cdot)$ and $\phi(\cdot )$ be respectively the cdf and pdf of the standard normal distribution. Let $x_n(u)$ be the solution of the equation $F_n(x)=\Phi(u)$ for $x$ in terms of $u$ and
\begin{eqnarray*}
\underline{l}_n(u)&=&n^{1/2}(\exp(u^2/n)-1)^{1/2}\\
\overline{l}_n(u)&=&n^{1/2}[\exp(u^2/ (n-1/2) )-1]^{1/2}.
\end{eqnarray*}
Then for all $u>0$ and $n>1/2$
\begin{equation*}
    \underline{l}_n(u)\leq x_n(u)\leq\overline{l}_n(u).
\end{equation*}
\end{lemma}

\section{Proofs}
\noindent\textbf{Proof of Proposition 1:} In Proposition 2 of Fan, Han \& Gu (2012), we can show that  with probability 1 $$
    \Var(p_0^{-1} V(t) |W_1,\cdots,W_k) = O(p^{-\delta}).
$$
This implies that
\begin{equation*}
\Big|\frac{1}{p_0} V(t)-\frac{1}{p_0}\sum_{i\in\{\text{true \ nulls}\}}P(P_i\leq t|W_1,\cdots,W_k)\Big|=O_p(p^{-\delta/2}).
\end{equation*}
By (C1), the desired conclusion follows.

\begin{lemma}
Let $q_t=F^{-1}_n(t)$ and $z_t=\Phi^{-1}(t)$ $(0<t<1)$, then $|q_t-z_t|<C_t/n$ where $C_t$ is a constant with respect to $n$.
\end{lemma}
\textbf{Proof of Lemma 4:} Note that $q_t=-q_{1-t}$ and $z_t=-z_{1-t}$. We only need to prove the inequality holds when $0.5<t<1$. Let $u=z_t$. Since $q_t>u>0$, we only need to show $\overline{l}_n(u)-u\leq C_t/n$ in light of Lemma 3. By Taylor expansion, we have
\begin{equation*}
\exp(u^2/ (n-1/2) ) -1= u^2/ (n-1/2) + h,
\end{equation*}
where  $h=u^4\exp(x^{\star})/[2(n-1/2)^2]$ and $0\leq x^{\star}\leq u^2/n$.  It is easy to see that $h \leq C_u/n^2$ for some constant $C_u$, independent of $n$. Therefore,
\begin{eqnarray*}
\overline{l}_n(u)-u&=&\sqrt{n} \sqrt{ u^2/({n-1/2}) + h} - u \\
         & \leq & \sqrt{n} \sqrt{ u^2/({n-1/2}) + C_u/n^2} - u.
\end{eqnarray*}
The last can easily be shown to be bounded by $C'_u/n$ for some positive constant $C'_u$.  The conclusion thus follows.

\textbf{Proof of Theorem 2:} To prove the first result in Theorem 2, by condition (C1),  it is sufficient to show that
\begin{equation*}
\Big|p_0^{-1}V(t)-p_0^{-1}\sum_{i\in\{\text{true \ nulls}\}} \big[\Phi(a_i(z_{t/2}+\eta_i))+\Phi(a_i(z_{t/2}-\eta_i))\big]\Big|=O_p(p^{-\delta/2})+O_p(n^{-1/2}).
\end{equation*}
To prove this, it suffices to show
\begin{equation}\label{j4}
\Big|p_0^{-1}\sum_{i\in\{\text{true \ nulls}\}}\bI(P_i\leq t | \bW)-p_0^{-1}\sum_{i\in\{\text{true \ nulls}\}}P(P_i\leq t|\bW)\Big|=O_p(p^{-\delta/2})+O_p(n^{-1/2}),
\end{equation}
and that
\begin{equation}\label{j5}
\Big|p_0^{-1}\sum_{i\in\{\text{true \ nulls}\}}P(P_i\leq t| \bW)-p_0^{-1}\sum_{i\in\{\text{true \ nulls}\}}[\Phi(a_i(z_{t/2}+\bb_i^T\bW))+\Phi(a_i(z_{t/2}-\bb_i^T\bW))]\Big|=O(n^{-1}).
\end{equation}

To prove (\ref{j4}), it is sufficient to show that
\begin{equation}\label{j6}
\Var\Big(p_0^{-1}\sum_{i\in\{\text{true \ nulls}\}}\bI(P_i\leq t|\bW)\Big)=O(p^{-\delta})+O(n^{-1}).
\end{equation}
The left hand side of (\ref{j6}) is
\begin{equation*}
p_0^{-2}\sum_{i\in\{\text{true \ nulls}\}}\Var(\bI(P_i\leq t| \bW))+p_0^{-2}\sum_{i,j\in\{\text{true \ nulls}\},i\neq j}\Cov(\bI(P_i\leq t|\bW), \bI(P_j\leq t| \bW)).
\end{equation*}
Since $\Var(\bI(P_i\leq t|\bW)\leq 1/4$, the first term above is $O(p^{-1})$.  For the second term, we have
\begin{eqnarray}\label{j7}
&&\Cov(\bI(P_i\leq t| \bW), \bI(P_j\leq t | \bW))\nonumber\\
&=&P(|T_i|\leq-q_{t/2}, |T_j|\leq-q_{t/2}|\bW)-P(|T_i|\leq-q_{t/2}|\bW)P(|T_j|<-q_{t/2}|\bW)
\end{eqnarray}

Let $\bV^{1/2}$ be a $p\times p$ diagonal matrix with diagonal elements $\{\sqrt{v_i}\}_{i=1}^p$. From the definition of the $\{T_i\}$ statistics, they have the following representation (it also admits a Bayesian interpretation of $t$-distribution):
\begin{eqnarray*}
(T_1,\cdots, T_p)^T\big|\{V_i=v_i\}_{i=1}^p&\sim& N_p(\bV\bmu, \bV^{1/2}\bSigma\bV^{1/2}),\\
V_i&\sim& InverseGamma(\frac{n-1}{2},\frac{n-1}{2})\quad\quad i=1,\cdots,p.
\end{eqnarray*}
In the above, the marginal distribution of $V_i \sim 1/\sqrt{\chi_{n-1}^2/(n-1)}$ is an inverse Gamma with degrees of freedom $((n-1)/2, (n-1)/2)$. When $n\rightarrow\infty$, $E(V_i)\rightarrow1$ and $\Var(V_i)\rightarrow0$. Therefore, $T_i$ converges to the limiting random variable $Z_i$. However, the joint distribution of $(V_1,\cdots,V_p)$, which depends on $\bSigma$, is very complicated because of the dependency among these random variables. Fortunately, thanks to the dominated convergence theorem, in the following proof, we do not need the explicit expression for this joint distribution. Since we only need to calculate the joint probability of bivariate case under the null hypothesis, the representation is even simpler. For each pair $(i,j)\in\{\text{true \ nulls}\}$, relating to equation (4) in the paper, we have
\begin{eqnarray*}
T_i|V_i=v_i,V_j=v_j&=&\sqrt{v_i}Z_i=\sqrt{v_i}(\bb_i^T\bW+K_i)\\
T_j|V_i=v_i,V_j=v_j&=&\sqrt{v_j}Z_j=\sqrt{v_j}(\bb_j^T\bW+K_j)\\
V_i \mbox{ and } V_j &\sim& InverseGamma(\frac{n-1}{2},\frac{n-1}{2})\\
(V_i,V_j)&\sim& f(v_i,v_j).
\end{eqnarray*}
Let $c_{1,i}=a_i(-q_{t/2}/\sqrt{v_i}-\bb_i^T\bW)$, $c_{2,i}=a_i(q_{t/2}/\sqrt{v_i}-\bb_i^T\bW)$, $c_{1,j}=a_j(-q_{t/2}/\sqrt{v_j}-\bb_j^T\bW)$, and $c_{2,j}=a_j(q_{t/2}/\sqrt{v_j}-\bb_j^T\bW)$. Then in the first term of (\ref{j7}), we can write
\begin{eqnarray}\label{j10}
&&P(q_{t/2}\leq T_i\leq-q_{t/2}, q_{t/2}\leq T_j\leq -q_{t/2}|\bW)\\
&=&\int_0^{\infty}\int_0^{\infty}P(c_{2,i}/a_i\leq K_i\leq c_{1,i}/a_i, c_{2,j}/a_j\leq K_j\leq c_{1,j}/a_j|\bW, v_i,v_j )f(v_i,v_j)dv_idv_j. \nonumber
\end{eqnarray}
Following the similar argument in the proof of Proposition 2 in Fan, Han \& Gu (2012), the integrand function is the joint cdf of bivariate normal random variables and can be expressed as
\begin{eqnarray}\label{j8}
&&P(c_{2,i}/a_i\leq K_i\leq c_{1,i}/a_i, c_{2,j}/a_j\leq K_j\leq c_{1,j}/a_j|\bW, v_i,v_j )\\
&=&\int_{-\infty}^{\infty}\big[\Phi(\frac{(\rho_{ij}^k)^{1/2}z+c_{1,i}}{(1-\rho_{ij}^k)^{1/2}})-\Phi(\frac{(\rho_{ij}^k)^{1/2}+c_{2,i}}{(1-\rho_{ij}^k)^{1/2}})\big]\big[\Phi(\frac{(\rho_{ij}^k)^{1/2}z+c_{1,j}}{(1-\rho_{ij}^k)^{1/2}})-\Phi(\frac{(\rho_{ij}^k)^{1/2}z+c_{2,j}}{(1-\rho_{ij}^k)^{1/2}})\big]\phi(z)dz,\nonumber
\end{eqnarray}
where $\rho_{ij}^k$ is the correlation of $K_i$ and $K_j$, and without loss of generality, we assume $\rho_{ij}^k>0$ here. For negative $\rho_{ij}^k$, we can obtain similar results. Let $cov_{ij}^k$ denote the covariance of $K_i$ and $K_j$, and let $b_{ij}^k=(1-\|\bb_i\|^2)^{1/2}(1-\|\bb_j\|^2)^{1/2}$, then similar to Fan, Han \& Gu (2012), for each $\Phi(\cdot)$, we apply Taylor expansion with respect to $(cov_{ij}^k)^{1/2}$, (\ref{j8}) can be written as
\begin{equation}\label{j9}
\big[\Phi(c_{1,i})-\Phi(c_{2,i})\big]\big[\Phi(c_{1,j})-\Phi(c_{2,j})\big]+\big(\phi(c_{1,i})-\phi(c_{2,i})\big)\big(\phi(c_{1,j})-\phi(c_{2,j})\big)(b_{ij}^k)^{-1}cov_{ij}^k+O(|cov_{ij}^k|^{3/2}).
\end{equation}
In (\ref{j9}), for each $\Phi(\cdot)$, we apply the second order Taylor expansion with respect to $(v_i-1)$ and $(v_j-1)$ since the inverse gamma random variable will concentrate around 1 as $n$ increases. For example,
\begin{equation}\label{n1}
\Phi(c_{1,i})=\Phi\big(a_i(-q_{t/2}-\bb_i^T\bW)\big)+\frac{1}{2}\phi\big(a_i(-q_{t/2}-\bb_i^T\bW)\big)a_iq_{t/2}(v_i-1)+H_i(c^*)(c^*)^2
\end{equation}
for some $c^*\in(0,v_i-1)$ if $v_i>1$ and $c^*\in(v_i-1,0)$ if $v_i<1$, where $H_i(\cdot)$ is the second derivative of $\Phi(c_{1,i})$ with respect to $(v_i-1)$. By the fact that $\exp(-x)\leq k!/x^k$ for any nonnegative integer number $k$, it is easy to show that $H_i(\cdot)$ is uniformly bounded on the set of $v_i$ with measure 1. For $\Phi(c_{2,i})$, $\Phi(c_{1,j})$ and $\Phi(c_{2,j})$, we have similar results.  Apply the Mean Value theorem to $\phi(c_{1,i})$, $\phi(c_{2,i})$, $\phi(c_{1,j})$ and $\phi(c_{2,j})$, we can also obtain similar results.

In (\ref{j7}), we can show that
\begin{equation}\label{n2}
P(|T_l|\leq-q_{t/2}|\bW)=\int_0^{\infty}[\Phi(c_{1,l})-\Phi(c_{2,l})]f(v_l)dv_l
\end{equation}
for the index $l=i, j$. Next we will evaluate the covariance between $\bI(P_i\leq t|\bW)$ and $\bI(P_j\leq t|\bW)$. Since $V_i$ follows the InverseGamma$((n-1)/2, (n-1)/2)$, we have
\begin{equation*}
EV_i=\frac{n-1}{n-3}, \quad \Var(V_i)=\frac{2(n-1)^2}{(n-3)^2(n-5)}, \quad E(V_i-1)^4=O(n^{-1}).
\end{equation*}
By Cauchy-Schwartz inequality, it is not difficult to show that $E|(V_i-1)(V_j-1)|=O(n^{-1})$, $E[|V_i-1|(V_j-1)^2]=O(n^{-1})$ and $E(V_i-1)^2(V_j-1)^2=O(n^{-1})$. Combining (\ref{j10}), (\ref{j8}), (\ref{j9}) with the above expressions for $\Phi(\cdot)$ and $\phi(\cdot)$, we have
\begin{eqnarray}\label{j11}
&&P( |T_i|\leq-q_{t/2}, |T_j|\leq-q_{t/2}|\bW)-P(|T_i|\leq-q_{t/2}|\bW)P(|T_j|\leq-q_{t/2}|\bW)\nonumber\\
&=&O(n^{-1})+\Big\{\big[\phi(a_i(-q_{t/2}-\bb_i^T\bW))-\phi(a_i(q_{t/2}-\bb_i^T\bW)\big]\big[\phi(a_j(-q_{t/2}-\bb_j^T\bW))
    \nonumber \\
 & & -\phi(a_j(q_{t/2}-\bb_j^T\bW))\big]a_ia_j + O(n^{-1/2})\Big\}\cov_{ij}^k+O(|\cov_{ij}^k|^{3/2}).
\end{eqnarray}
Note that the coefficient before $\cov_{ij}^k$ in (\ref{j11}) is uniformly bounded. Therefore, in (\ref{j7}),
\begin{equation*}
\Cov(\bI(P_i\leq t|\bW), \bI(P_j\leq t|\bW))=O(|\cov_{ij}^k|)+O(|\cov_{ij}^k|^{3/2})+O(n^{-1}).
\end{equation*}
By the Cauchy-Schwartz inequality and condition (C0), we have
\begin{equation*}
p^{-2}\sum_{i,j}|\cov_{ij}^k|\leq p^{-1}[\sum_{i,j}(\cov_{ij}^k)^2]^{1/2}=p^{-1}[\sum_{j=k+1}^p\lambda_j^2]^{1/2}=O(p^{-\delta}).
\end{equation*}
Also we have $|\cov_{ij}^k|^{3/2}<|\cov_{ij}^k|$. Therefore, we can conclude that
\begin{equation*}
\Var\Big(p_0^{-1}\sum_{i\in\{\text{true \ nulls}\}}\bI(P_i\leq t|\bW)\Big)=O(p^{-\delta})+O(n^{-1}).
\end{equation*}
This establishes (\ref{j4}).

We now prove (\ref{j5}). Similar to the discussion for (\ref{n1}) and (\ref{n2}), we can show that
\begin{equation*}
P(P_i\leq t|\bW)=\Phi(a_i(q_{t/2}+\bb_i^T\bW))+\Phi(a_i(q_{t/2}-\bb_i^T\bW))+O(n^{-1}).
\end{equation*}
From Lemma 4 in Supplementary Materials, we know $z_{t/2}=q_{t/2}+\Delta$, where $0<\Delta\leq C_t/n$ and $C_t$ is a constant, independent of $n$. By the mean value theorem,
\begin{eqnarray*}
\Phi(a_i(q_{t/2}+\eta_i))&=&\Phi(a_i(z_{t/2}+\eta_i))-\Delta a_i\phi(x_{i1}^{\star})\\
\Phi(a_i(q_{t/2}-\eta_i))&=&\Phi(a_i(z_{t/2}-\eta_i))-\Delta a_i\phi(x_{i2}^{\star})
\end{eqnarray*}
where $a_i(z_{t/2}+\eta_i)-\Delta a_i<x_{i1}^{\star}<a_i(z_{t/2}+\eta_i)$ and $a_i(z_{t/2}-\eta_i)-\Delta a_i<x_{i2}^{\star}<a_i(z_{t/2}-\eta_i)$. Thus, (\ref{j5}) can be expressed as
\begin{eqnarray}\label{h6}
\Big|p_0^{-1}\Delta\sum_{i\in\{true \ null\}}a_i[\phi(x_{i1}^{\star})+\phi(x_{i2}^{\star})]\Big|=O(n^{-1}),
\end{eqnarray}
as $a_i[\phi(x_{i1}^{\star})+\phi(x_{i2}^{\star})]$ is uniformly bounded for every $i$. This completes the proof of the first result.

For the second result, define an infeasible estimator
\begin{equation*}
\widetilde{\bW}_2=(\bB^T\bB)^{-1}\bB^T\bT.
\end{equation*}
Denote $\FDP_2(t)$ as the estimator in equation (7) with using the infeasible estimator $\widetilde{\bW}_2$. Then
\begin{equation*}
\widehat{\FDP}_{U,G}(t)-\FDP_A(t)=[\widehat{\FDP}_{U,G}(t)-\FDP_2(t)]+[\FDP_2(t)-\FDP_1(t)]+[\FDP_1(t)-\FDP_A(t)],
\end{equation*}
where $\FDP_1(t)$ is defined in the proof of Theorem 1.

Similar to the proof of Theorem 1, we can show that
\begin{equation*}
\big|\widehat{\FDP}_{U,G}(t)-\FDP_2(t)\big|\leq(R(t))^{-1}\Big\{C_1\sum_{h=1}^k\big[|\widehat{\lambda}_h-\lambda_h|+\lambda_h\|\widehat{\bgamma}_h-\bgamma_h\|\big]+C_2p^{1/2}\sum_{h=1}^k\|\widehat{\bgamma}_h-\bgamma_h\|\|\bT\|\Big\}
\end{equation*}
for some positive constants $C_1$ and $C_2$.

As shown in the proof for the first result of Theorem 2, $\bT=\bV\bZ$ where $\bV=\diag\{\sqrt{V_i}\}$ and
\begin{equation*}
V_i\sim InverseGamma(\frac{n-1}{2},\frac{n-1}{2})\quad\quad i=1,\cdots,p,
\end{equation*}
independent of $Z_i$.  Using this representation, we have
\begin{equation*}
    E \|\bT\|^2 = \sum_{i=1}^p E V_i Z_i^2 = \frac{n-1}{n-3} ( \|\bmu^{\star}\|^2 + p ).
\end{equation*}
This implies that
\begin{equation*}
\|\bT\|=O_p(\|\bmu^{\star}\|+p^{1/2}).
\end{equation*}

Similarly, we can also show that
\begin{equation*}
|\FDP_2(t)-\FDP_1(t)|\leq (R(t))^{-1}C_3p^{1/2}\big\|\sum_{h=1}^k\bgamma_h\bgamma_h^T\big\|\|\bT-\bZ\|.
\end{equation*}
Stochastically, we have
\begin{equation*}
E \|\bT-\bZ\|^2 \leq   \sum_{i=1}^p E(\sqrt{V_i} - 1)^2 Z_i^2  = E(\sqrt{V_i} - 1)^2 ( \|\bmu^{\star}\|^2 + p ).
\end{equation*}
Using $E(\sqrt{V_i}-1)^2=O(n^{-1})$, it follows that
$$
   \|\bT-\bZ\|=O_p\{ n^{-1/2} (\|\bmu^{\star}\| + p^{1/2}) \}.
$$

For $|\FDP_1(t)-\FDP_A(t)|$, we have shown the result in the proof of Theorem 1. Combining all the results above, the proof is now complete.

\section{Additional Simulation and Data Results}

\subsection{Comparison with the benchmark with known covariance.}
\begin{table}
\caption{\label{Tab12} Empirical mean absolute error between true $\FDP(t)$ and $\widehat{\FDP}_A(t)$ for known covariance and $\widehat{\FDP}_{POET}(t)$ for unknown covariance. Results are in percent. }
\centering
\begin{tabular}{l|rccc}
\hline\hline
     &Sample Size   & LAD & LS  & SCAD  \\
\hline
    Model 1, Known Covariance &$n=50$ & 2.78    & 2.69  & 3.56   \\
          &$n=100$ &2.73   & 2.61 & 3.93\\
          &$n=200$ &2.63    & 2.46  & 4.87   \\
\hline
    Model 1, Unknown Covariance &$n=50$ & 4.26   & 4.06  & 4.54 \\
     & $n=100$ & 3.73 & 3.63   & 4.80  \\
     &$n=200$ & 3.26  & 3.10   & 5.12 \\
\hline\hline
     Model 2, Known Covariance   &$n=50$ & 3.22    & 3.37  & 3.14 \\
        &$n=100$ & 3.53  & 3.41 & 3.46\\
        &$n=200$ & 3.99   & 3.89  & 4.78  \\
\hline
      Model 2, Unknown Covariance  &$n=50$ & 4.63   & 4.56  & 4.73 \\
        &$n=100$ & 4.38    & 4.31   & 4.36  \\
        &$n=200$ & 4.50  & 4.35   & 6.17 \\
\hline\hline
\end{tabular}
\end{table}

We first compare the realized $\FDP(t)$ values with $\widehat{\FDP}_A(t)$ given in (7) and $\widehat{\FDP}_{\text{POET}}(t)$ to evaluate the performance of our POET-PFA procedure. Note that $\widehat{\FDP}_A(t)$ is constructed based on a known covariance matrix $\bSigma$ and is used as a benchmark for $\widehat{\FDP}_{\text{POET}}(t)$. We apply three different estimators for the realized but unknown factors: least absolute deviation estimator (LAD)  (10), least squares estimator (LS) (11) and smoothly clipped absolute deviation estimator (SCAD) (8).  Fan, Han \& Gu (2012) has theoretically and numerically shown that $\widehat{\FDP}_A(t)$ performs well. The performance of LAD and SCAD under unknown dependence can be better illustrated through an apparent factor model structure. Therefore, we only present the results corresponding to Models 1 \& 2. For other models considered in section 3.1, LAD and SCAD might not be very effective. We have the simulation results for $n=50, 100, 200$, but due to the space limit, we will only present the results for $n=50$. Figures 1 and 2 correspond to strict factor model and approximate factor model respectively. They show clearly that both $\widehat{\FDP}_A(t)$ and  $\widehat{\FDP}_{\text{POET}}(t)$ approximate $\FDP(t)$ very well.  In addition, they demonstrate that $\widehat{\FDP}_{\text{POET}}(t)$ performs comparably with but slightly inferior to $\widehat{\FDP}_A(t)$.  This shows that the price paid to estimate the unknown covariance matrix is limited.  Table 1 provides additional evidence to support the statement, in which we compute the mean absolute error between the approximated FDP and the true FDP.
\begin{figure}[h!!!]
\begin{center}
\scalebox{0.62}{\includegraphics{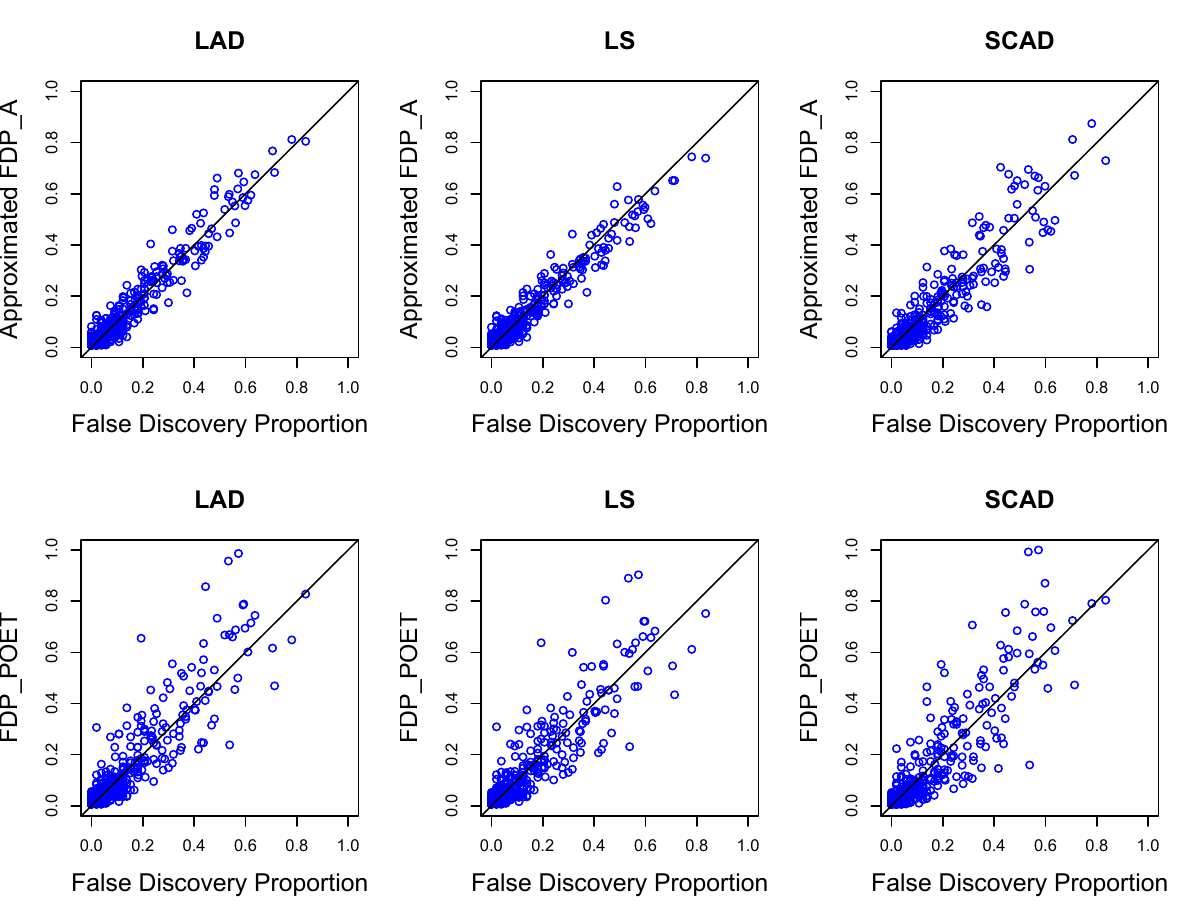}}
\end{center}
\vspace{-0.3cm}
\caption{Comparison of realized values of False Discovery Proportion with $\widehat{\text{FDP}}_A(t)$ and $\widehat{\text{FDP}}_{\text{POET}}(t)$ for Model 1. }\label{Fig1}
\end{figure}
\begin{figure}[h!!!]
\begin{center}
\scalebox{0.62}{\includegraphics{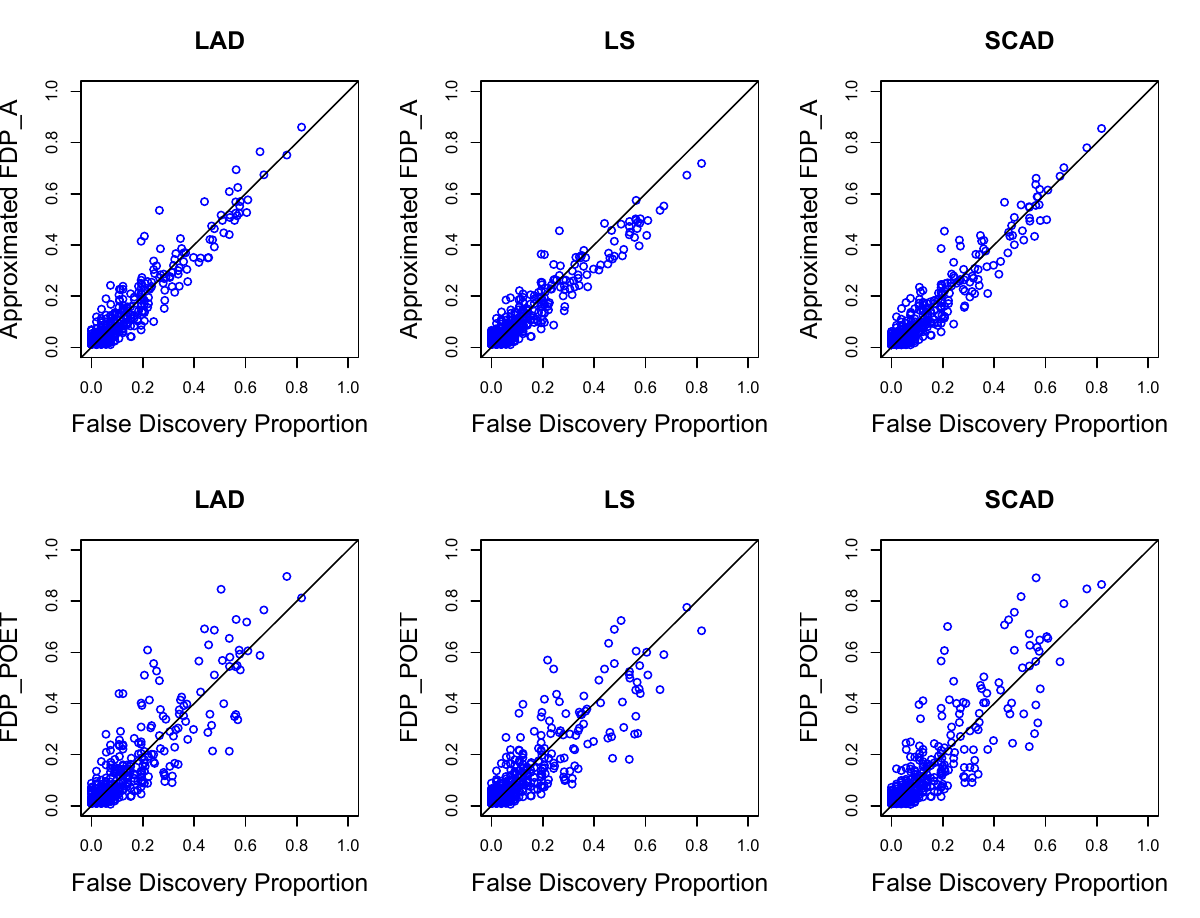}}
\end{center}
\vspace{-0.3cm}
\caption{Comparison of realized values of False Discovery Proportion with $\widehat{\text{FDP}}_A(t)$ and $\widehat{\text{FDP}}_{\text{POET}}(t)$ for Model 2.}
\end{figure}

\subsection{Comparison with other methods}
We further compare POET-PFA with other methods under different signal strength. The detailed results are shown in Tables 2 \& 3. Overall, POET-PFA is still the best in terms of producing smaller mean absolute error. It is worth mentioning that HF-PFA is very competitive and outperforms under several model settings with certain sample size. 

Figures 3 \& 4 illustrate the performance of POET-PFA with least squares estimator compared with Efron, FAMT and FAMT-PFA under Models 3-8. Although the dependence structures vary across the model settings, our POET-PFA still captures the trend of the true FDP. 

\begin{table}
\caption{\label{Tab3} Empirical mean absolute error between true $\FDP(t)$ and $\widehat{\FDP}(t)$. The nonzero $\mu_i=0.8$. The results are in percent. }
\begin{tabular}{cccccccc}
\hline\hline
       & POET-PFA & Efron  & FAMT  & FAMT-PFA  & HF-PFA & SS-PFA  & LW-PFA  \\
\hline\hline
Model 1   &    &     &     &  &    &   &     \\
     $n=50$   & 4.56 & 19.92   & 10.87   & 6.20   & 5.24    & 5.93   & 5.17     \\
     $n=100$  &3.88   & 19.82    & 11.56    & 5.89    & 6.80    & 4.90    & 4.56   \\
     $n=200$  & 3.89  & 19.54   & 11.99    & 5.82    & 4.83    & 4.48     & 4.29   \\
\hline
Model 2    &   &   &   &  &     &    &   \\
        $n=50$  & 5.00   & 18.94    &  11.33     & 6.54      & 5.47       &  6.45    & 5.89   \\
        $n=100$ & 4.04   & 17.66      & 10.13      & 5.05     & 4.61       & 5.22      &  5.04  \\
        $n=200$ & 4.01    & 17.58   & 10.39     & 4.98     & 4.59     & 4.56     & 4.50   \\
\hline
Model 3  &      &     &       &    &     &     &      \\
     $n=50$     & 7.55   & 16.82   &  15.67   & 9.43    & 5.64   & 7.27    & 6.47    \\
     $n=100$   & 4.36    & 14.34   & 12.15   & 6.46     & 3.98    & 4.86    & 4.48  \\
     $n=200$   &  3.84   & 14.26    & 13.04   & 6.09    & 4.76   & 4.66    & 4.49   \\
\hline
Model 4  &      &     &       &    &    &    &    \\
     $n=50$     &  5.19   & 19.03   & 11.81  &  7.00  & 5.53   & 7.84    & 7.60    \\
     $n=100$   &  4.02   & 19.39   & 10.06   & 6.09  & 6.16    & 4.94   & 4.98    \\
     $n=200$   & 3.63    & 19.80    & 10.02   & 6.14   & 4.34   & 4.17    & 4.32     \\
\hline
Model 5  &      &     &       &    &     &     &     \\
     $n=50$     &  5.46     & 10.28       & 10.22     & 5.54    & 5.67    & 7.02   & 5.73  \\
     $n=100$   &  5.58     & 11.28       & 10.86     & 5.46    & 6.56  & 6.32    & 5.92   \\
     $n=200$   &  5.15     & 10.52       & 10.53     & 4.87   & 7.13  & 5.56     & 5.39    \\     
\hline
Model 6  &      &     &       &     &     &     &      \\
     $n=50$     &  5.33     & 10.89       & 10.19       &  5.60    & 5.54    & 6.56    & 5.33  \\
     $n=100$   &  4.24    & 9.76       & 9.55       & 4.37   &  5.08    & 4.94    &  4.24   \\
     $n=200$   &  4.14    & 9.47       & 9.49       & 4.13     & 3.83   & 4.50    & 4.14    \\   
\hline
Model 7  &      &     &       &    &    &    &       \\
     $n=50$     & 4.46    & 10.61      & 6.02     & 4.73    & 5.01   & 6.02   & 4.51   \\
     $n=100$   & 4.11      &  10.17      &  6.39      & 4.50   & 5.44    & 5.07   &  4.21   \\
     $n=200$   & 4.11      &  10.43      &  6.69      & 4.71    & 6.12   & 4.78    & 4.21   \\   
\hline
Model 8  &      &     &       &      &     &     &    \\
     $n=50$     & 4.25      &  10.77      &  5.51    &  4.47   & 5.16    & 5.97    & 4.43    \\
     $n=100$   & 4.44      &  11.85      &  6.81     & 4.92   & 4.49    & 5.96    & 4.82    \\
     $n=200$   & 4.12      &  11.44      & 6.55       & 4.81   & 3.70    & 4.73    & 4.28    \\       
\hline\hline       
\end{tabular}
\end{table}

\begin{table}
\caption{\label{Tab4} Empirical mean absolute error between true $\FDP(t)$ and $\widehat{\FDP}(t)$. The nonzero $\mu_i=1.2$. The results are in percent.}
\begin{tabular}{cccccccc}
\hline\hline
       & POET-PFA & Efron  & FAMT  & FAMT-PFA  & HF-PFA & SS-PFA  & LW-PFA  \\
\hline\hline
Model 1    &   &   &   &  &     &    &   \\
        $n=50$  & 4.49  & 19.12  & 11.25  & 5.59  & 4.78  & 6.82  & 6.00 \\
        $n=100$ & 3.68  & 19.56  & 10.23 & 4.83 & 4.95   & 4.88    & 4.54\\
        $n=200$ &3.50   & 19.22  & 9.95 & 4.51   & 2.91   & 3.99    & 3.85 \\
\hline
Model 2   &    &     &     &  &    &   &     \\
     $n=50$   & 4.81   & 18.41     & 11.15   & 5.59  & 4.99   & 7.09    & 6.54     \\
     $n=100$  & 4.31    & 18.57    & 11.03     & 5.85   & 5.08    & 5.24    & 5.15     \\
     $n=200$  & 3.75    & 18.27     & 11.01    & 5.49   & 4.55   & 4.29    & 4.24   \\
\hline
Model 3  &      &     &       &    &     &     &      \\
     $n=50$     &   5.13    & 14.31    & 12.85     & 6.81     & 5.61    &  8.14   & 6.75  \\
     $n=100$   &  4.25     & 14.22     & 12.12    & 5.88     & 5.56    & 6.00    & 5.59   \\
     $n=200$   &  3.28     & 14.17     & 11.78    & 5.46     & 4.90    & 4.21    & 4.07   \\
\hline
Model 4  &      &     &       &    &    &    &    \\
     $n=50$     &  4.81     & 19.90       & 11.11       & 6.29  & 5.48   & 7.66     & 8.07     \\
     $n=100$   &  3.79     & 19.03       & 11.18       & 6.73  & 4.54    &  4.81  & 4.92    \\
     $n=200$   &   3.54    & 19.02       & 10.51       & 6.24 &  4.97  & 4.20    & 4.26     \\
\hline
Model 5  &      &     &       &    &     &     &     \\
     $n=50$     & 5.83      & 10.98       & 11.15       & 5.81     & 6.19    & 7.40   & 5.90  \\
     $n=100$   & 5.55      & 10.78       & 10.91       & 5.24     & 5.96   & 6.40    & 5.63   \\
     $n=200$   & 5.53      & 10.39       & 11.42       & 5.00   &6.07   & 5.98     & 5.64    \\     
\hline
Model 6  &      &     &       &     &     &     &      \\
     $n=50$     & 4.39       & 9.65       & 9.28       & 4.67     & 4.38    & 5.81    & 4.39  \\
     $n=100$   & 4.10      & 9.41       & 9.24       & 4.25   &  5.84    & 4.86    & 4.10     \\
     $n=200$   &  4.33     & 9.95      & 10.11    & 4.35      & 4.16    & 4.85    & 4.34    \\   
\hline
Model 7  &      &     &       &    &    &    &       \\
     $n=50$     & 4.39      & 10.45       & 6.32       & 4.90    & 4.97   & 6.60    & 4.74    \\
     $n=100$   & 4.24      & 10.02       &  6.35      & 4.74   & 4.40    & 5.46   & 4.47    \\
     $n=200$   &  4.22     & 10.00       &  6.57      & 4.76    & 4.46   & 4.75    & 4.28   \\   
\hline
Model 8  &      &     &       &      &     &     &    \\
     $n=50$     &  4.44     & 12.01    &5.97     & 4.62   & 4.36   & 6.58    & 4.84    \\
     $n=100$   & 4.24     & 11.67        & 6.20        & 4.68   & 4.52    & 5.32    & 4.47     \\
     $n=200$   & 4.25      &  10.91     & 7.05       & 5.00   & 5.30    & 4.82    & 4.40    \\       
\hline\hline       
\end{tabular}
\end{table}

\begin{figure}[h!!!]
\begin{center}
\scalebox{0.56}{\includegraphics{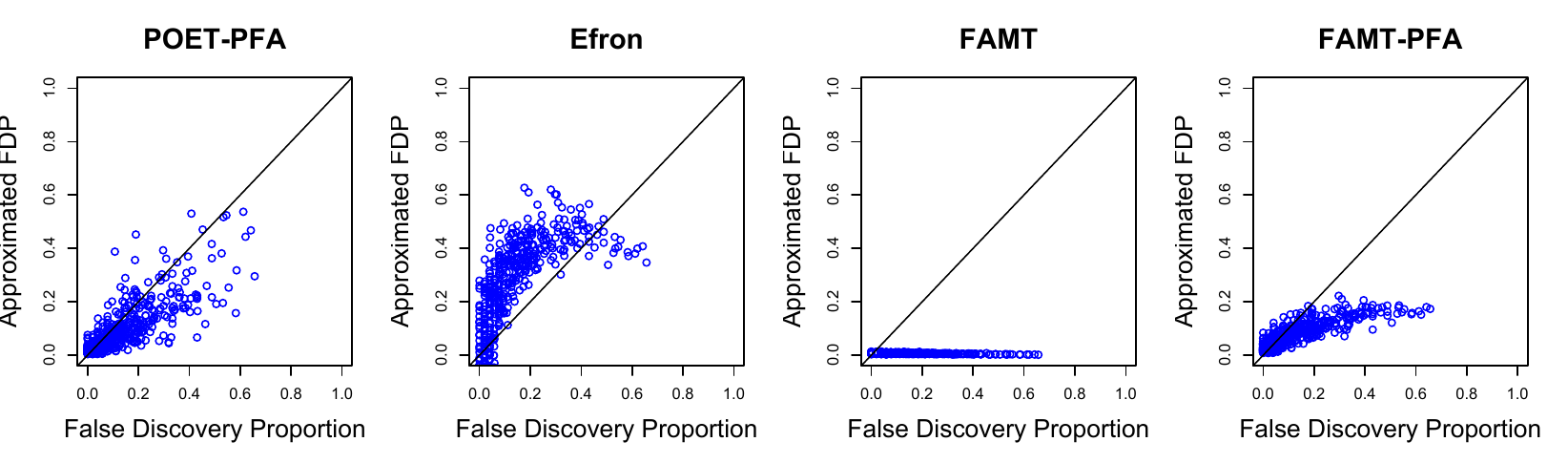}}
\scalebox{0.56}{\includegraphics{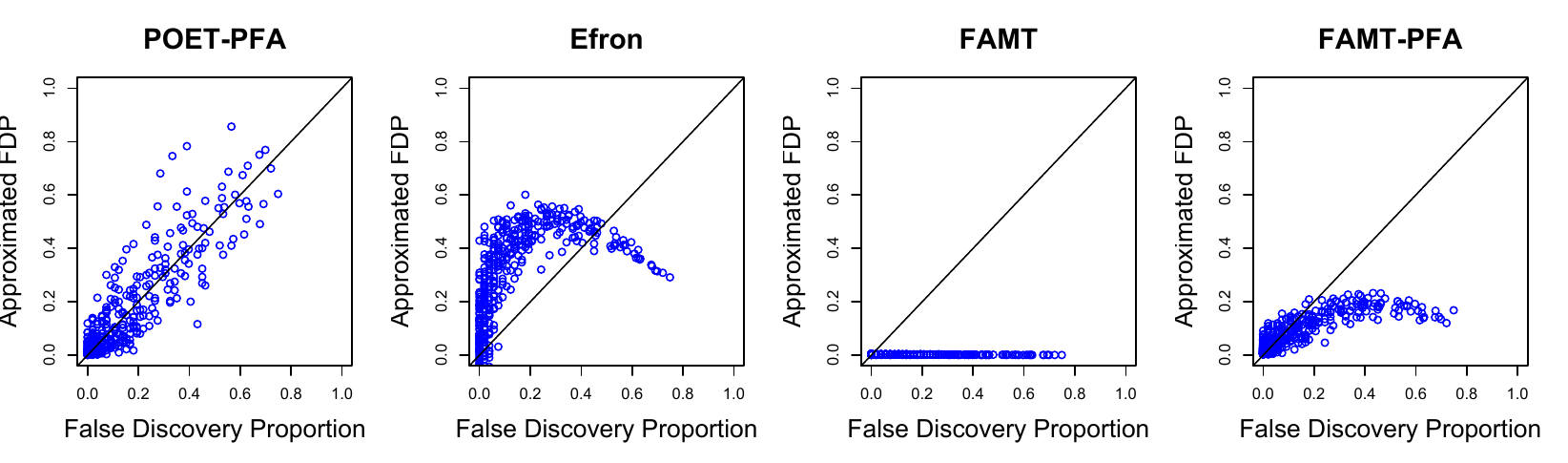}}
\scalebox{0.56}{\includegraphics{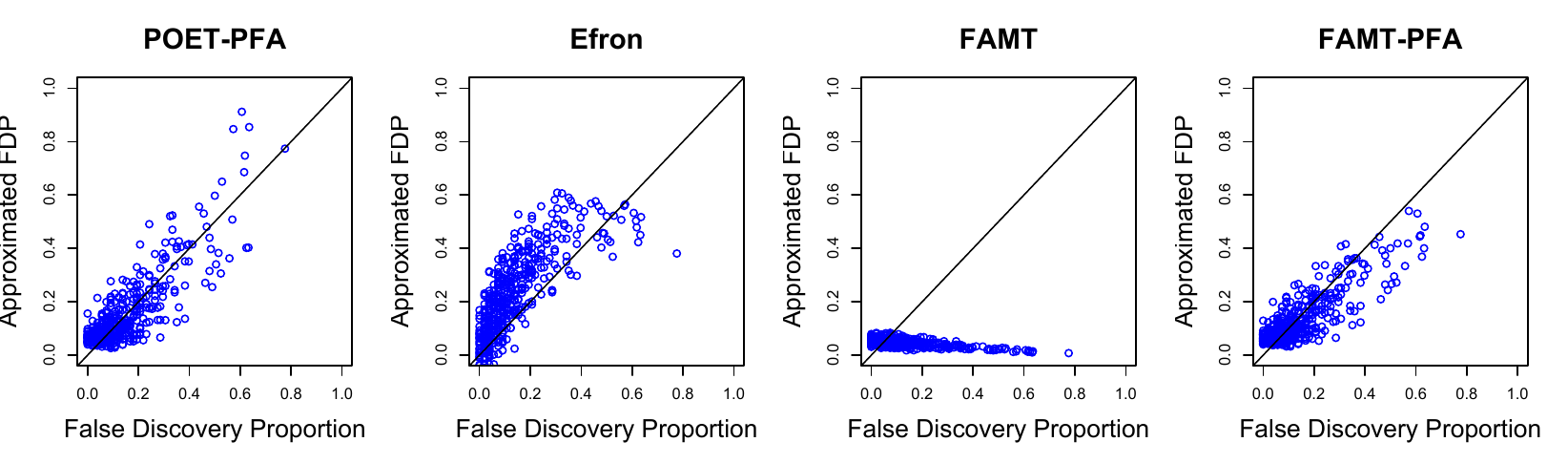}}
\scalebox{0.56}{\includegraphics{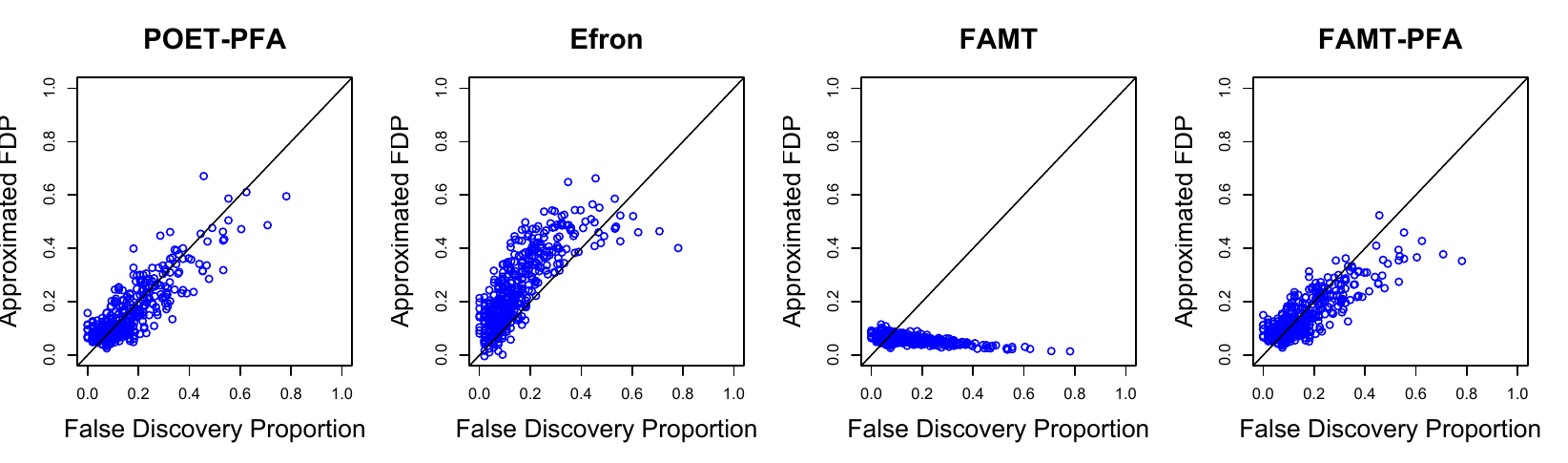}}
\end{center}
\vspace{-0.3cm}
\caption{Comparison of realized values of False Discovery Proportion with $\widehat{\text{FDP}}_{\text{POET}}(t)$ involving least-squares estimation, Efron (2007) estimator, FAMT, and FAMT-PFA.  From top to bottom, the panels correspond to Models 3-6. $n=50$. Nonzero $\mu_i=1$. }
\end{figure}
\begin{figure}[h!!!]
\begin{center}
\scalebox{0.56}{\includegraphics{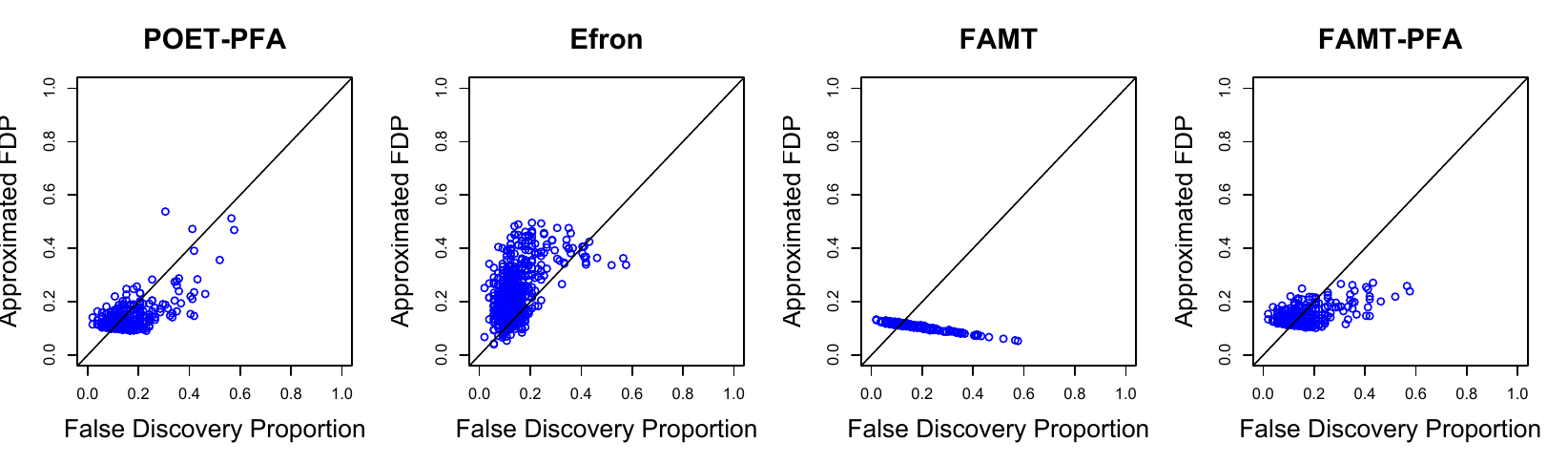}}
\scalebox{0.56}{\includegraphics{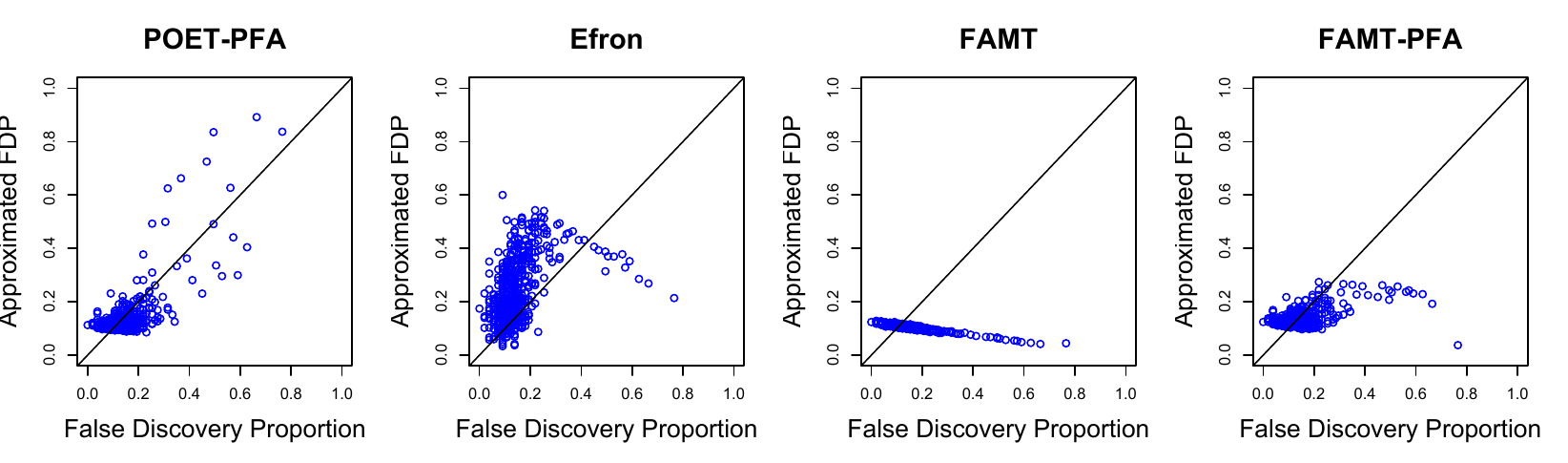}}
\end{center}
\vspace{-0.3cm}
\caption{Comparison of realized values of False Discovery Proportion with $\widehat{\text{FDP}}_{\text{POET}}(t)$ involving least-squares estimation, Efron (2007) estimator, FAMT, and FAMT-PFA.  From top to bottom, the panels correspond to Models 7 \& 8. $n=50$. Nonzero $\mu_i=1$. }
\end{figure}

\subsection{Data Analysis}
In Figure 5, we summarize the relationship of approximated FDP and number of total rejections. Compared with Figure 2 of the main paper, the approximated FDP tends to be smaller with the same amount of total rejections. The 40 most significantly differentially expressed genes are listed in Tables 4 \& 5 for the fixed threshold method and the dependence adjusted method. 
\begin{figure}[h!!!]
\begin{center}
\scalebox{0.65}{\includegraphics{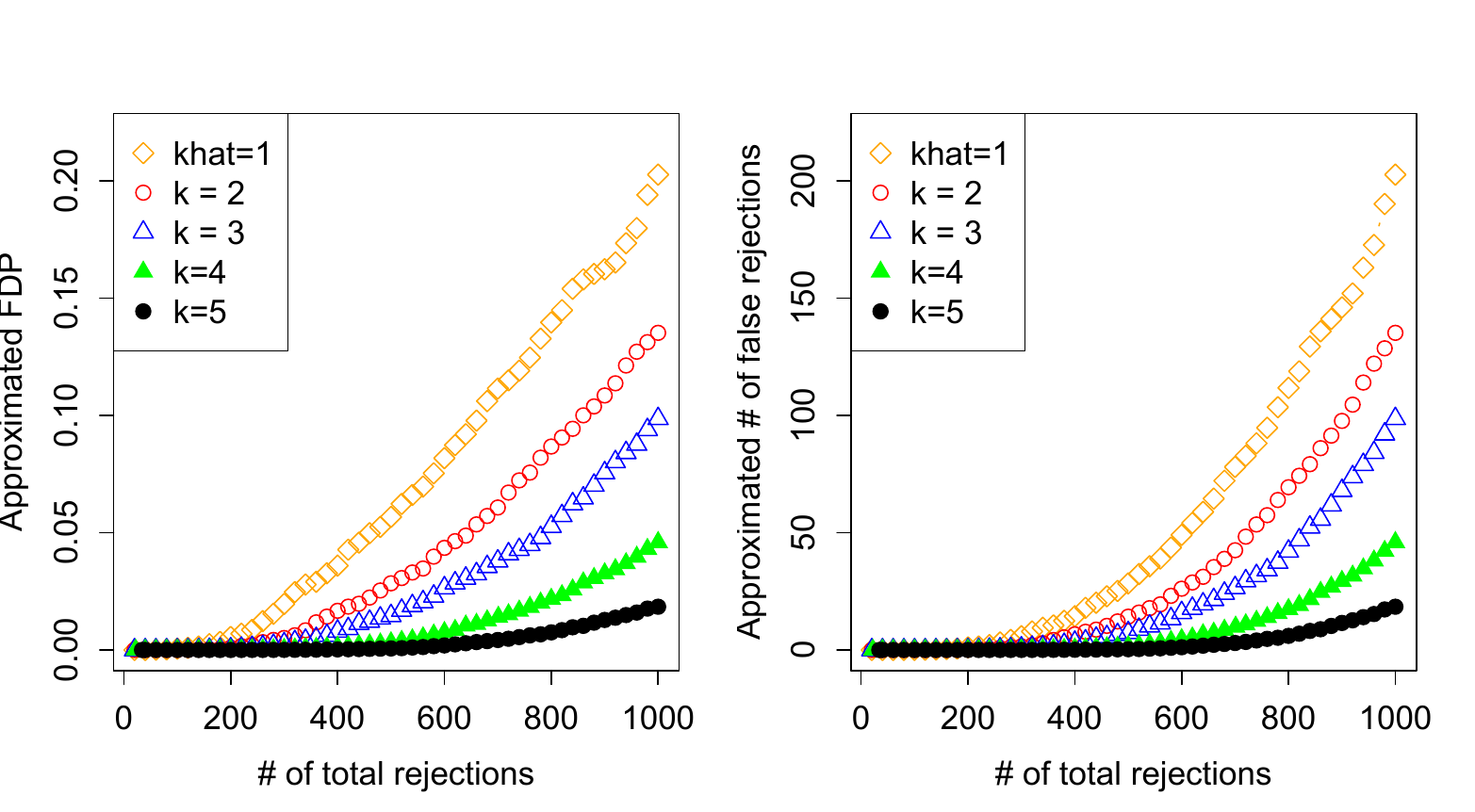}}
\end{center}
\vspace{-0.3cm}
\caption{The approximated false discovery proportion and the approximated number of false discoveries as functions of
the number of total discoveries for $p=3226$ genes, where the estimated $k$ is 1 compared with other choices $k=2,3,4,5$, using dependence-adjusted procedure. }\label{Fig100}
\end{figure}

\begin{table}
\caption{\label{Tab100}40 most significantly differentially expressed genes that can discriminate breast cancers with BRCA1
mutations from those with BRCA2 mutations. The approximated FDP is approximately $0.02\%$ under approximate factor model with 1 factor, providing strong evidence for our selection.}
\small
\begin{tabular}{ll}
\hline\hline
Clone ID & UniGene Title \\
\hline
26184 & phosphofructokinase, platelet\\
810057 & cold shock domain protein A\\
46182 & CTP synthase\\
813280 & adenylosuccinate lyase\\
950682 & phosphofructokinase, platelet\\
840702 & SELENOPHOSPHATE SYNTHETASE ; Human selenium donor protein\\
784830 & D123 gene product\\
841617 & Human mRNA for ornithine decarboxylase antizyme, ORF 1 and ORF 2\\
563444 & forkhead box F1\\
711680 & zinc finger protein, subfamily 1A, 1 (Ikaros)\\
949932 & nuclease sensitive element binding protein 1\\
75009 & EphB4\\
566887 & chromobox homolog 3 (Drosophila HP1 gamma)\\
841641 & cyclin D1 (PRAD1: parathyroid adenomatosis 1)\\
809981 & glutathione peroxidase 4 (phospholipid hydroperoxidase)\\
236055 & DKFZP564M2423 protein\\
293977 & ESTs, Weakly similar to putative [C.elegans]\\
295831 & ESTs, Highly similar to CGI-26 protein [H.sapiens]\\
236129 & Homo sapiens mRNA; cDNA DKFZp434B1935\\
247818 & ESTs\\
814270 & polymyositis/scleroderma autoantigen 1 (75kD)\\
130895 & ESTs\\
548957 & general transcription factor II, i, pseudogene 1\\
212198 & tumor protein p53-binding protein, 2\\
293104 & phytanoyl-CoA hydroxylase (Refsum disease)\\
82991 & phosphodiesterase I/nucleotide pyrophosphatase 1\\
32790 & mutS (E. coli) homolog 2 (colon cancer, nonpolyposis type 1)\\
291057 & cyclin-dependent kinase inhibitor 2C (p18, inhibits CDK4)\\
344109 & proliferating cell nuclear antigen\\
366647 & butyrate response factor 1 (EGF-response factor 1)\\
366824 & cyclin-dependent kinase 4\\
471918 & intercellular adhesion molecule 2\\
136769 & TATA box binding protein (TBP)\\
23014 & mitogen-activated protein kinase 1\\
26184 & phosphofructokinase, platelet\\
29054 & ARP1 (actin-related protein 1, yeast) homolog A (centractin alpha)\\
36775 & hydroxyacyl-Coenzyme A dehydrogenase\\
42888 & interleukin enhancer binding factor 2, 45kD\\
45840 & splicing factor, arginine/serine-rich 4\\
51209 & protein phosphatase 1, catalytic subunit, beta isoform\\
\hline\hline
\end{tabular}
\end{table}

\begin{table}
\caption{\label{Tab200}40 most significantly differentially expressed genes that can discriminate breast cancers with BRCA1
mutations from those with BRCA2 mutations under dependence-adjusted procedure. The approximated FDP is approximately $0.0032\%$ under approximate factor model with 1 factor, providing strong evidence for our selection.}
\small
\begin{tabular}{ll}
\hline\hline
Clone ID & UniGene Title \\
\hline
    26184 & phosphofructokinase, platelet \\
  752631 & fibroblast growth factor receptor 3 (achondroplasia, thanatophoric dwarfism)\\
  810057 & cold shock domain protein A\\
  813280 & adenylosuccinate lyase\\ 
  714106 & plasminogen activator, urokinase \\
  950682 & phosphofructokinase, platelet\\
  784830 & D123 gene product \\
  841617 & Human mRNA for ornithine decarboxylase antizyme, ORF 1 and ORF 2 \\
  711680  & zinc finger protein, subfamily 1A, 1 (Ikaros)\\
  784360  & echinoderm microtubule-associated protein-like \\
  949932  &nuclease sensitive element binding protein 1 \\
   75009   &EphB4\\
  784224  &fibroblast growth factor receptor 4 \\
  566887  &chromobox homolog 3 (Drosophila HP1 gamma)\\
 841641   &cyclin D1 (PRAD1: parathyroid adenomatosis 1) \\
 205049   &ESTs, Weakly similar to heat shock protein 27 [H.sapiens] \\
 768561   &small inducible cytokine A2 (monocyte chemotactic protein 1, homologous to mouse Sig-j \\
 809981   &glutathione peroxidase 4 (phospholipid hydroperoxidase)\\
 236055   &DKFZP564M2423 protein\\
 293977   &ESTs, Weakly similar to putative [C.elegans]\\
 295831   &ESTs, Highly similar to CGI-26 protein [H.sapiens]\\
 236129   &Homo sapiens mRNA; cDNA DKFZp434B1935 (from clone DKFZp434B1935) \\
 247818   &ESTs \\
 243360   &ESTs, Moderately similar to cytoplasmic dynein intermediate chain 1 [H.sapiens] \\
 814270   &polymyositis/scleroderma autoantigen 1 (75kD)\\
 140635   &ESTs\\
 548957   &general transcription factor II, i, pseudogene 1\\
 212198   &tumor protein p53-binding protein, 2 \\
 293104   &phytanoyl-CoA hydroxylase (Refsum disease) \\
  82991    &phosphodiesterase I/nucleotide pyrophosphatase 1 (homologous to mouse Ly-41 antigen \\
  32790    &mutS (E. coli) homolog 2 (colon cancer, nonpolyposis type 1) \\
 291057  &cyclin-dependent kinase inhibitor 2C (p18, inhibits CDK4) \\
 366647  &butyrate response factor 1 (EGF-response factor 1) \\
 366824  &cyclin-dependent kinase 4 \\
 361692  &sarcoma amplified sequence \\
  26184   &phosphofructokinase, platelet\\
  29054   &ARP1 (actin-related protein 1, yeast) homolog A (centractin alpha) \\
  36775   &hydroxyacyl-Coenzyme A dehydrogenase/3-ketoacyl-Coenzyme A thiolase/enoyl-Coenzy\\
  42888   &interleukin enhancer binding factor 2, 45kD\\
  51209   &protein phosphatase 1, catalytic subunit, beta isoform \\
\hline\hline
\end{tabular}
\end{table}

\end{document}